\documentclass{aa}
\usepackage{hyperref}
\usepackage[varg]{txfonts} 
\usepackage{natbib}
\usepackage{ctable}
\usepackage{graphicx}
\usepackage{braket}
\usepackage{comment}
\usepackage{ulem}
\begin{document}

\title{Galaxy cluster mass accretion rates from IllustrisTNG} 

\author{Michele Pizzardo \inst{\ref{1}}\thanks{\email{michele.pizzardo@smu.ca}}
\and Margaret J. Geller \inst{\ref{2}}
\and Scott J. Kenyon \inst{\ref{2}}
\and Ivana Damjanov \inst{\ref{1}}
\and Antonaldo Diaferio \inst{\ref{3},\ref{4}}
}

\institute{\label{1}Department of Astronomy and Physics, Saint Mary's University, 923 Robie Street, Halifax, NS-B3H3C3, Canada
\and \label{2}Smithsonian Astrophysical Observatory, 60 Garden Street, Cambridge, MA-02138, USA
\and \label{3}Dipartimento di Fisica, Universit\`a di Torino, via P. Giuria 1,  I-10125 Torino, Italy
\and \label{4}Istituto Nazionale di Fisica Nucleare (INFN), Sezione di Torino, via P. Giuria 1,  I-10125 Torino, Italy
} 

\date{Received date / Accepted date}

\abstract
{ We use simulated cluster member galaxies from Illustris TNG300-1 to develop a technique for measuring the galaxy cluster mass accretion rate (MAR) that can be applied directly to observations. We analyze 1318  IllustrisTNG clusters of galaxies with $M_{200c}>10^{14}$M$_\odot$ and $0.01\leq z \leq 1.04$.  The  MAR we derive is the ratio between the mass of a spherical shell located in the infall region and the time for the infalling shell to accrete onto the virialized region of the cluster.\\ 
At fixed redshift, an $\sim 1$ order of magnitude increase in $M_{200c}$ results in a comparable increase in MAR. At fixed mass, the MAR increases by a factor of $\sim 5$ from $z=0.01$ to $z=1.04$. The MAR estimates derived from the caustic technique  are unbiased and lie within 20\% of the MARs based on the true mass profiles. This agreement is crucial for observational derivation of the MAR. The IllustrisTNG results are also consistent with (i) previous merger tree approaches based on N-body dark matter only simulations and with (ii) previously determined  MARs of real clusters based on the caustic method. Future  spectroscopic and photometric surveys will provide MARs of enormous cluster samples with mass profiles derived from both spectroscopy and weak lensing. Combined with future larger volume hydrodynamical simulations that extend to higher redshift, the MAR promises important insights into evolution of massive systems of galaxies. 
}


\maketitle

\section{Introduction}\label{sec:introduction} 

In the standard $\Lambda$CDM cosmological model, where gravity follows general relativity in a flat spacetime filled with a positive cosmological constant ($\Lambda$), cold collisionless dark matter (CDM), and ordinary baryonic matter, clusters of galaxies form hierarchically  by progressive accretion of matter onto density peaks \citep[e.g.,][]{press1974formation,white1978,BBKS86,bower1991,laceyCole93,sheth2002,zhang2008,corasaniti2011,desimone2011,achitouv2014,musso2018}.
The outskirts of clusters of galaxies are potentially powerful testbeds for this accretion predicted by structure formation models \citep{diaferio2004outskirts,reiprich2013outskirts,Diemer2014,lau2015mass,walker2019physics,deBoni2016,rost2021threehundred}. 

The mass accretion rate (MAR)  probes  the outer (infall) region of galaxy clusters because the accretion draws material from radii $\gtrsim R_{200c}$.  $R_{200c}$, a common proxy for the virial radius, is the radius enclosing an average mass density 200 times the critical density of the Universe at the appropriate redshift. For radii $\gtrsim R_{200c}$ clusters are not in dynamical equilibrium \citep{ludlow2009,Diemer2014,More2015,bakels2020,pizzardo2020}.
The MAR is naturally linked to the splashback radius, the average location of the first apocenter of infalling material \citep{adhikari2014,Diemer2014,More2015}. This radius is located at $\sim (1-2)R_{200c}$ and decreases with increasing MAR, mass, and redshift \citep{Diemer2017sparta2,oneil21}.
N-body simulations show that MARs correlate with other properties of cluster halos, including concentration \citep{Wechsler_2002,Tasitsiomi_2004,zhang2004,giocoli2012,Ludlow2013}, shape \citep{kasun2005shapes,allgood2006shape,bett2007spin,ragone2010relation}, degree of internal relaxation \citep{power2011}, and fraction of substructure \citep{gao2004sub,vandenbosch2005,Ludlow2013}. 

Many theoretical studies investigate the MAR in $\Lambda$CDM cosmologies. The first approaches to computing the MAR in $\Lambda$CDM  cosmologies were analytic models based on the extended Press-Schechter (EPS) formalism or on Monte-Carlo generated merger trees \citep{bower1991,laceyCole93}. More recent approaches build merger trees from N-body dark matter only simulations \citep{mcbride2009,Fakhouri2010,Diemer2017sparta2,Diemer18} or from semi-analytical models calibrated with N-body simulations \citep{vandenbosch14,Correa15b}.  
These models imply that halos with mass $\gtrsim 10^{14}$M$_\odot$ accrete $\sim 30\%-50\%$ of their mass over the redshift range $z\sim 0.5$ to $z\sim 0$. The MAR is correlated with both halo mass and redshift. Because we observe a cluster at a single redshift and thus cannot  measure its history directly, the merger tree approaches are not directly applicable to  cluster observations.

Soon, spectroscopic and photometric missions will provide huge samples of clusters with both dense spectroscopy and weak-lensing maps extending to large cluster-centric radius. The cluster samples will also cover redshifts $\lesssim 2$. Both weak-lensing \citep{bartelmann2010gravitational,hoekstra2013masses,Umetsu20essay} and  the caustic technique \citep{Diaferio1997,Diaferio99,Serra2011,Pizzardo23} will allow  estimates of cluster mass profiles without assuming dynamical equilibrium.  

The use of IllustrisTNG \citep{Pillepich18,Springel18,Nelson19} enables direct application of the caustic technique to track the MAR for galaxy clusters with $z \lesssim 1$. In contrast with a MAR recipe based on purely N-body simulations \citep{pizzardo2020}, IllustrisTNG enables the use of galaxies as tracers of the dynamical evolution of galaxy clusters. Following \citet{pizzardo2020} we define the MAR as the ratio between the mass of an infalling shell and the infall time. \citet{pizzardo2020} and \citet{Pizzardo2022} compute the MAR based on mass profiles determined from the caustic technique applied to  dense spectroscopic surveys of the infall regions of observed  galaxy clusters  \citep{Rines2006CIRS,Rines2013HeCS,sohn2021hectomap,sohn2021cluster}. The results are consistent with $\Lambda$CDM predictions.

\citet{Pizzardo23} use IllustrisTNG to calibrate a statistical platform for application of the caustic technique. Their approach is based on  mock galaxy cluster member catalogs. The caustic technique calibrated by IllustrisTNG returns the true cluster mass profile within  10\% in the radial range $(0.6-4.2)R_{200c}$ and redshift range $0.01-1.04$. 

We build on the approach of \citet{Pizzardo23} to develop an IllustrisTNG recipe for computing the cluster MAR. The recipe is based on simulated galaxies rather than dark matter particles in contrast with previous approaches. We demonstrate that the caustic mass profiles based on galaxy mock catalogs \citep{Pizzardo23} yield reliable estimates of the true MARs. They are  also consistent with  estimates from previous theoretical and observational investigations \citep{mcbride2009,Fakhouri2010,vandenbosch14,Correa15a,Correa15b,Diemer2017sparta2,Diemer18,pizzardo2020,Pizzardo2022}.

Sect. \ref{sec:sample} describes the IllustrisTNG  cluster sample. We derive the average radial velocity profiles, fundamental ingredients for the computation of the MAR. Sect. \ref{sec:recipe} describes the approach  to estimating the MAR. In Sect. \ref{sec:mshelltinf} we derive the mass of the infalling shell and the infall time. Sect. \ref{sec:results} compares the true MARs with the caustic MARs. Sect. \ref{sec:discussion} compares these MAR results with previous models  and observations.  We  also compare galaxy and dark matter MARs. Finally we outline future challenges and prospects. We conclude in Sect. \ref{sec:conclusion}.

\section{Cluster sample, mass and velocity profiles}
\label{sec:sample}

Basic inputs to the mass accretion rate include the cluster mass and radial velocity profiles.
A large sample of clusters \citep{Pizzardo23} from the IllustrisTNG simulations \citep{Pillepich18,Springel18,Nelson19} is the basis for the determination of the true 3D  and projected caustic cumulative mass profiles. As a basis for computing the mass accretion rate,  we compute the average  cluster radial velocity profile at each redshift. This profile has a characteristic minimum that ultimately determines the accretion rate.

Sect. \ref{subsec:tng} describes the sample of clusters extracted from IllustrisTNG  by \citet{Pizzardo23}. We summarize the derivation of the  true and caustic mass profiles for each cluster.  Sect. \ref{subsec:vrad} describes the determination of the radial velocity profiles.

\subsection{Cluster sample and mass profiles}
\label{subsec:tng}

\citet{Pizzardo23} extract cluster samples from the TNG300-1 run of the IllustrisTNG simulations \citep{Pillepich18,Springel18,Nelson19}, a set of gravo-magnetohydrodynamical simulations based on the $\Lambda$CDM model. Table \ref{table:tng_det} lists the cosmological parameters of the simulations. 
TNG300-1 is the baryonic run with the highest resolution among the runs with the largest simulated volumes. The simulation has a comoving box size of $302.6$~Mpc. TNG300-1 contains $2500^3$ dark matter particles with mass $m_{\rm DM} = 5.88 \times 10^{7}~\rm{M_{\odot}}$ and the same number of gas cells with average mass $m_b = 1.10\times 10^{7}~\rm{M_{\odot}} $.

\begin{table}[htbp]
\begin{center}
\caption{\label{table:tng_det} Cosmological parameters for IllustrisTNG.}
\begin{tabular}{llc}
\hline
\hline
Parameter & Description & Value \\
\hline
$\Omega_{\Lambda 0}$ & cosmological constant & 0.6911 \\
$\Omega_{m0}$ & total matter density & 0.3089 \\
$\Omega_{b0}$ & baryonic matter density & 0.0486 \\
$H_0$ & Hubble constant & $67.74$~km~s$^{-1}$~Mpc$^{-1}$ \\
$\sigma_8$ & power spectrum norm. & 0.8159 \\
$n_s$ & power specrum index & 0.9667 \\
\hline
 \end{tabular}
 \end{center}
\end{table}  

\citet{Pizzardo23} use group catalogues compiled by the IllustrisTNG Collaboration to extract all of the Friends-of-Friends (FoF) groups in TNG300-1 with $M_{200c}^{3D} > 10^{14}$M$_\odot$. There are 1697 clusters in the 11 redshift bins: $z=0.01, 0.11, 0.21, 0.31, 0.42, 0.52, 0.62, 0.73, 0.82, 0.92$, and $1.04$. For $\sim 22\%$ of the clusters sparse sampling, rich foreground or background, or the presence of significant substructures preclude robust application of the caustic technique \citep{Pizzardo23}. We remove these clusters from the sample. 

Table \ref{table:3dinfo} describes the remaining 1318 clusters we analyze here.  The Table includes the number of clusters in each redshift bin, the median and the interquartile range of their masses $M_{200c}^{3D}$, and the minimum and maximum $M_{200c}^{3D}$ at each redshift.

\begin{table}[htbp]
\begin{center}
\begin{small}
\caption{\label{table:3dinfo} Cluster samples from Illustris TNG300-1.}
\begin{tabular}{ccccc}
\hline
\hline
 $z$ & No. of & Median $M_{200c}^{3D}$ & 50\% range & (min-max) $M_{200c}^{3D}$  \\
     & clusters & $\left[10^{14}~\text{M}_{\odot }\right]$  & $\left[10^{14}~\text{M}_{\odot }\right]$ & $\left[10^{14}~\text{M}_{\odot }\right]$  \\
\hline
 & & \\
0.01 & 224 & 1.64 & 1.24 -- 2.48 & 1.00 -- 15.0 \\
0.11 & 212 & 1.59 & 1.26 -- 2.39 & 1.01 -- 12.6 \\
0.21 & 181 & 1.48 & 1.22 -- 2.25 & 1.01 -- 9.65 \\
0.31 & 157 & 1.49 & 1.24 -- 2.15 & 1.02 -- 8.24 \\
0.42 & 134 & 1.43 & 1.20 -- 1.91 & 1.00 -- 7.68 \\
0.52 & 109 & 1.39 & 1.20 -- 1.91 & 1.00 -- 7.54 \\
0.62 & 97 & 1.42 & 1.16 -- 1.83 & 1.00 -- 5.29 \\
0.73 & 70 & 1.32 & 1.15 -- 1.74 & 1.01 -- 4.34 \\
0.82 & 57 & 1.38 & 1.12 -- 1.78 & 1.01 -- 4.47 \\
0.92 & 44 & 1.33 & 1.11 -- 1.88 & 1.00 -- 4.30 \\
1.04 & 33 & 1.43 & 1.15 -- 1.91 & 1.02 -- 4.37 \\
\hline
 \end{tabular}
 \end{small}
 \end{center}
\end{table}

Our goal is assessment  of the caustic technique \citep{Diaferio1997,Diaferio99,Serra2011,Pizzardo23} as a basis for reliable estimates of the true MAR. Estimation of the MAR requires robust knowledge of the cluster mass at large cluster-centric distances, $\gtrsim 2R_{200c}$, where virial equilibrium does not hold (Sects. \ref{sec:introduction} and \ref{sec:recipe}).   
In the extended radial range $(0.6-4.2)R_{200}$ the caustic technique returns an unbiased estimate of the mass with better than 10\% accuracy and with a relative uncertainty of $23\%$ provided that the velocity field of the cluster outer region is sufficiently well sampled \citep{Pizzardo23}. The caustic technique is independent of equilibrium assumptions.

The true 3D  and caustic MARs  rest on determination of the true and caustic mass profiles for each cluster in the sample (Table \ref{table:3dinfo}). \citet{Pizzardo23} compute the true cumulative mass profile (from now on, the ``true mass profile'') for each cluster from the 3D distribution of matter extracted from raw snapshots. These profiles include all matter species: dark matter, gas, stars, and black holes. \citet{Pizzardo23} compute the true mass profile for each cluster, $M^{3D}(r)$, in 200 logarithmically spaced bins covering  the radial range $(0.1-10)R_{200c}^{3D}$. These profiles define $R_{200c}^{3D}$ and $M_{200c}^{3D}$ for each cluster.

The $r-v_{\rm los}$ diagram, the line-of-sight velocity relative to the cluster median redshift as a function of $r$, $v_{\rm esc, los}(r)$ \citep{Diaferio1997,Diaferio99}, is the basis for estimating the caustic cumulative mass profile (caustic mass profile hereafter). 
In the  $r-v_{\rm los}$  diagram, cluster members appear within a well-defined trumpet-shaped pattern. The amplitude of the pattern decreases as $r$ increases. The caustic technique locates the boundaries (the ``caustics'') of this pattern that discriminate between cluster members and the foreground/background. The caustic amplitude, $\mathcal{A}(r)$ is half of the vertical separation between the upper and the lower caustic at radius $r$. \citet{Diaferio1997} show that the caustic amplitude approximates the escape velocity from the cluster, $\mathcal{A}(r)\approx v_{\rm esc,los}(r)$. The square of the caustic amplitude, $\braket{v_{\rm esc,los}^2(r)}$, is then a measure of the gravitational potential.  

The caustic technique estimate of the mass profile is
\begin{equation}\label{eq:CT}
	GM (<r) = {\cal F}_\beta \int_{0}^{r} {\cal A}^2(R) \,{\rm d}R ~ ,
\end{equation}
where $G$ is the gravitational constant and ${\cal F}_\beta$ is a constant filling factor. \citet{Pizzardo23} use the Illustris TNG simulations to  calibrate the filling factor in the redshift range $0.01-1.04$. We use their calibration, $\mathcal{F}_\beta=0.41\pm 0.08$, throughout.

The $r-v_{\rm los}$ diagram is based on catalogues of simulated cluster galaxies that include the right ascension $\alpha$, declination $\delta$, and total redshift $z$ along the line of sight. 
\citet{Pizzardo23} associate a realistic galaxy mock redshift survey with each simulated cluster. These catalogues include contaminating background and foreground galaxies. \citet{Pizzardo23} build the catalogues by identifying galaxies with Subfind substructures that have stellar mass $ > 10^8~$M$_\odot$. This choice of mass  limit mimics observable galaxies and it assures  optimal performance of the caustic technique.
Finally, \citet{Pizzardo23} apply the caustic technique in the same unconstrained way to  all of the mock catalogues and obtain a single calibrated caustic mass profile for each cluster. These profiles define $R_{200c}^{C}$ and $M_{200c}^{C}$ for each cluster.

\subsection{Radial velocity profiles}
\label{subsec:vrad}

The average cluster galaxy velocity profile along the radial direction is fundamental to the evaluation of the mass accretion rate (Sect. \ref{sec:recipe}). We compute the set of individual cluster radial velocity profiles based on the comoving position of simulated galaxies with respect to the cluster center, $\mathbf{r}_{c,i}$, and the galaxy peculiar velocity, $\mathbf{v}_{p,i}$. From the 3D volume we select only  galaxies with cluster-centric distances $ < 10 R_{200c}^{3D}$. We compute the radial velocity of each galaxy: $v_{{\rm rad},i}=[{\bf v}_{p,i} + H(z_s)a(z_s){\bf r}_{c,i}]\cdot {\bf r}_{c,i}/r_{c,i}$, where $H(z_s)$ and $a(z_s)$ are the Hubble function and the scale factor at the redshift $z_s$ of the snapshot.  We compute the mean radial velocity profile based on the galaxies within 100 linearly spaced radial bins covering  the range $(0,10)R_{200c}^{3D}$.  

For each redshift snapshot, we compute a single average radial velocity profile by averaging  over all of the individual radial velocity profiles for clusters within the snapshot (Table \ref{table:3dinfo}). For each of the 100 radial bins of $r/R_{200c}^{3D}$, we compute the mean and the standard deviation of all of the velocities within all of the clusters velocities in each bin. For each redshift, we thus obtain from all the individual clusters a single average mean radial velocity profile with the dispersion around it.

\begin{figure*}
    \centering
    \includegraphics[scale=0.73]{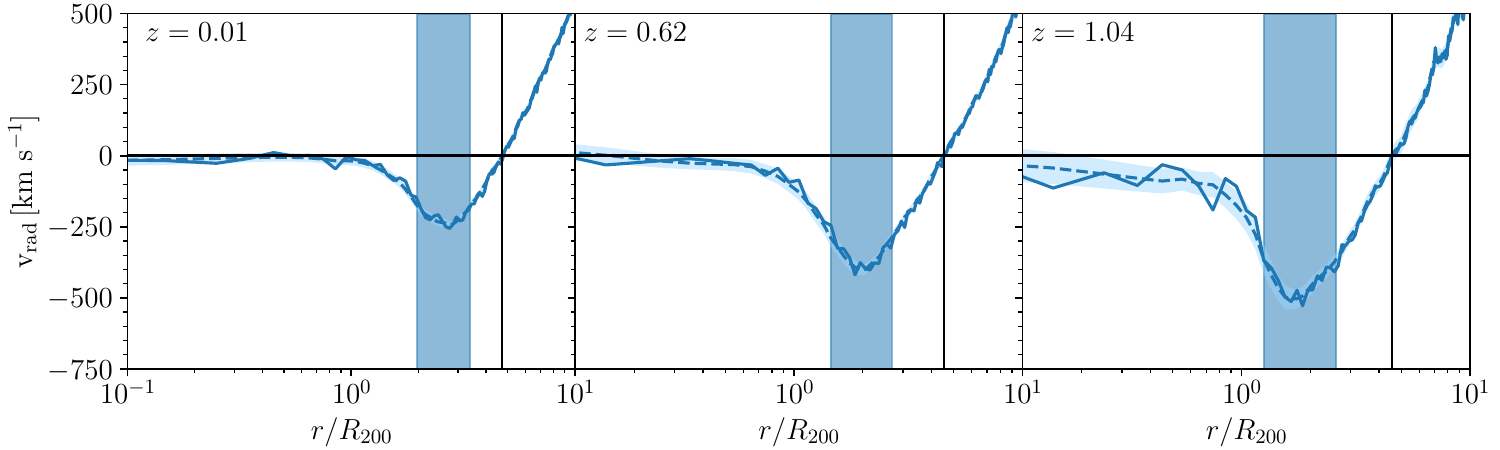}
    \caption{Galaxy radial velocity profiles, infalling shells, and turnaround radii.  The solid blue curves show the average galaxy radial velocity profiles for redshifts $z=0.01$, $z=0.62$, and $z=1.04$ (left, middle and right panels, respectively). Dashed blue curves show the Savitzky-Golay smoothed  profiles based on a ten-bin window \citep{Savitzky64}. The shadowed region shows the error in the smoothed profile. The vertical black line shows the average turnaround radius based on the velocity profiles. In each panel the  blue vertical band indicates the infalling shell with boundaries at cluster-centric radii where the velocity is $0.72 v_{min}$ (Sect. \ref{sec:mshelltinf}).}
    \label{fig:vrad}
\end{figure*}

The solid blue curves in Fig. \ref{fig:vrad} show mean radial velocity profiles at three example redshifts: $z=0.01$, $z=0.62$, and $z=1.04$ (in the left, middle, and right panels, respectively). We smooth over statistical fluctuations in the profile by applying a Savitsky-Golay filter with a 10 radial bin window \citep{Savitzky64}. The dash-dotted curves show the resulting smoothed profiles. The shaded light blue band shows the error in the corresponding smoothed profiles.

The curves show three clear regimes. Within $\sim 1 R_{200c}^{3D}$, the average radial velocity has a plateau at zero or small velocity. This  region is virialized, with no net infall. At larger distances from the cluster center, $\sim (1-4)R_{200c}^{3D}$, the infall region, the radial velocity is definitely non-zero and directed toward the  cluster center.  Finally, at distances $\gtrsim 4R_{200}$ the galaxy velocity is directed out of the cluster because the galaxies are still coupled to the universal Hubble flow.

Estimation of the MAR according to Eq. (\ref{eq:mar}) requires determination of the minimum of the radial velocity profile (see Sects. \ref{sec:recipe} and \ref{sec:mshelltinf}). At each redshift, we measure the minimum velocity of the smoothed average radial profile. The first three columns of Table \ref{table:rng_inf} list the minimum radial velocity, $v_{min}$, and its cluster-centric location, $R_{v_{min}}$, for each redshift bin. To compute the error in $R_{v_{min}}$, we bootstrap 1000 samples of clusters at each redshift (Table \ref{table:3dinfo}, columns 1 and 2).

Table \ref{table:3dinfo} shows that the median cluster mass generally decreases as the redshift increases because very massive systems are progressively less abundant at greater redshift. We check the impact of the changing distribution of  cluster masses on $R_{v_{min}}$. For each redshift, we build homogeneous samples that include only clusters with mass $M_{200c}^{3D}$ in the range $(1.02-4.30)\cdot 10^{14}$~M$_\sun$. The largest minimum and smallest maximum mass for the samples in the 11 redshift bins  set this range (see last column of Table \ref{table:3dinfo}). According to the Kolmogorov-Smirnov test, these clipped samples share  indistinguishable  mass distributions; the p-values are in the range $(0.4-0.9)$. For each redshift, we compute the average radial velocity profile for the clipped sample and  locate the turnaround radius. The turnaround radii for these samples are within $\lesssim 1.7\%$ of the turnaround radii of the full samples. They are also unbiased.
Hence the $R_{v_{min}}$ radii are insensitive to the difference among the distribution of cluster masses at different redshifts.

\begin{figure}
    \centering
    \includegraphics[width=\columnwidth]{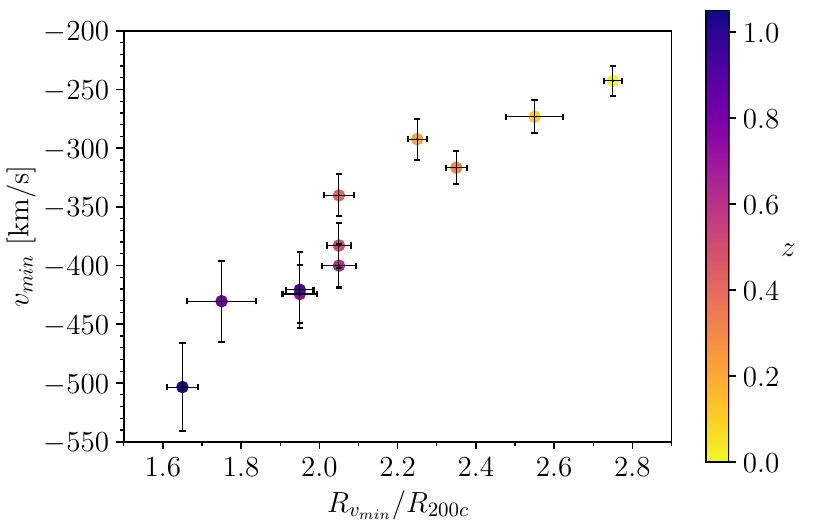}
    \caption{Minimum radial velocity as a function of the cluster-centric radius  of the minimum, colour coded by redshift.}
    \label{fig:vminvsrmin}
\end{figure}
Fig. \ref{fig:vminvsrmin} shows $v_{min}$ as a function of its $R_{v_{min}}$, colour coded by redshift. The minimum velocity and its cluster-centric radius are strongly correlated with redshift. From $z=0.01$ to $z=1.04$ the $v_{min}$ of clusters with approximately equal mass (Table \ref{table:3dinfo}) increases by $\sim 100\%$ in absolute value. The corresponding $R_{v_{min}}$ decreases by $\sim 40\%$. 

The behavior of $v_{min}$ as a function of redshift  is in qualitative agreement with the general picture of  hierarchical structure formation. For  equal mass clusters, systems at higher redshift form within higher overdensity peaks. Because of their denser environment, these clusters are denser within $\sim R_{200c}$. Thus more mass is confined within  smaller cluster-centric radii than for an equally  massive lower redshift cluster. The consequently deeper gravitational potential within $\sim R_{200c}$ at greater redshift is the source of the larger minimum infall velocity located at smaller cluster-centric radius. The corresponding mass accretion is also larger in these higher redshift, denser environments.  N-body simulations \citep[e.g.][]{vandenbosch02,mcbride2009,Fakhouri2010,vandenbosch14,Diemer2017sparta2} show that at fixed mass, the  accretion rate  at $z\sim 1$ exceeds the rate at  $z\sim 0$ by roughly an order of magnitude. 

\begin{table}[htbp]
\begin{center}
\caption{\label{table:rng_inf} Radial velocity minima and radii as a function of redshift.}
\begin{tabular}{ccccc}
\hline
\hline
 $z$ & $v_{min}$ & $R_{v_{min}}/R_{200c}$ & $v_{\rm inf}$ & Infalling shell  \\
     & [km~s$^{-1}$] & & [km~s$^{-1}$] &  $[R_{200c}]$ \\
\hline
 & & & & \\
0.01 & $-243\pm 13$ & $2.75\pm 0.02$ & $-214\pm 3$ & 1.97--3.40 \\
0.11 & $-273\pm 14$ & $2.55\pm 0.02$ & $-238\pm 4$ & 1.91--3.21 \\
0.21 & $-292\pm 18$ & $2.25\pm 0.05$ & $-257\pm 4$ & 1.70--3.07 \\
0.31 & $-317\pm 9$ & $2.35\pm 0.05$ & $-277\pm 4$ &  1.57--2.97 \\
0.42 & $-340\pm 18$ & $2.05\pm 0.06$ & $-300\pm 5$ &  1.57--2.86 \\
0.52 & $-382\pm 19$ & $2.05\pm 0.05$ & $-332\pm 5$ &  1.44--2.76 \\
0.62 & $-400\pm 19$ & $2.05\pm 0.07$ & $-344\pm 6$ &  1.45--2.68 \\
0.73 & $-424\pm 25$ & $1.95\pm 0.10$ & $-366\pm 7$ &  1.41--2.57 \\
0.82 & $-430\pm 35$ & $1.75\pm 0.10$ & $-372\pm 8$ &  1.33--2.61 \\
0.92 & $-421\pm 32$ & $1.95\pm 0.12$ & $-368\pm 8$ &  1.22--2.70 \\
1.04 & $-503\pm 37$ & $1.65\pm 0.15$ & $-432\pm 9$ &  1.25--2.60 \\
\hline
 \end{tabular}
 \end{center}
\end{table}

The dynamics within the infall region is independent of the turnaround radius where galaxies depart from the Hubble flow as a result of the cluster potential \citep{Gunn1972,silk1974,Schechter80}. We identify the turnaround radius from the smoothed radial velocity profiles. Starting from the largest cluster-centric distances, the turnaround radius is the first intersection between the profile and the axis $v_{\rm rad}=0$. The vertical solid lines in Fig. \ref{fig:vrad} show that the turnaround radii agree well with \citet{Meiksin86}. 
The turnaround radii are in the range $(4.52-4.78)R_{200c}$, $\sim (2-3)R_{200c}$ beyond $R_{v_{min}}$. In contrast with $R_{v_{min}}$, the turnaround radius decreases by only  $\sim 3.3\%$ as the redshift increases from $z=0.01$ to $z=1.04$. 

\section{Recipe for the estimation of the MAR}
\label{sec:recipe}

The average radial velocity profiles (Sect. \ref{subsec:vrad}, Fig. \ref{fig:vrad}) have clear minima at cluster-centric radii $\sim (2-3)R_{200c}$ identifying the region where clusters accrete new material. For an ideal, theoretical MAR (hereafter MAR$_t$), we first calculate the rate at which material falls through an infinitesimal shell at this minimum velocity. However, real and simulated systems have too few galaxies in this shell. Thus, we derive a second MAR in a larger volume that approximates MAR$_t$. We analyze the Illustris TNG300-1 simulations to relate the observational MAR to MAR$_t$.

Assuming spherical symmetry, the mass of an infinitesimal shell at a distance $r$ from the cluster center is dM$_{\rm shell}=4\pi \rho(r)r^2 {\rm d}r$, where d$r$ is the width of the shell and $\rho$ is the shell density. The time derivative of the shell mass is $\dot{M}={\text{dM}_{\rm shell}}/{{\rm d}t}$, where ${\rm d}t={\rm d}r/{\rm v}_{\rm rad}(r)$ is the time for infalling material with velocity $\rm v_{rad}(r)$ to cross d$r$. We define  a theoretical mass accretion rate MAR$_{\rm t} = \dot{M}(R_{v_{min}})$, where $R_{v_{min}}$ is the  minimum in the radial velocity profile (Sect. \ref{subsec:vrad}, Fig. \ref{fig:vrad}, and Table \ref{table:rng_inf}):
\begin{equation} \label{eq:martheo}
{\rm MAR}_{\rm t}=4\pi \rho(R_{v_{min}})R_{v_{min}}^2 v_{min}.
\end{equation}
IllustrisTNG provides an estimate of  ${\rm MAR}_{\rm t}$ for  a thin, finite shell where the radial velocity $\approx v_{min}$ is nearly constant acrosss the shell.

Observations can only provide the mass of an infalling shell with a finite width, $\Delta R \gg {\rm d}r$. 
We define the observational MAR of clusters of galaxies  as the accretion of matter within a spherical shell  onto the virialized region of the cluster:
\begin{equation}\label{eq:mar}
    \text{MAR}= \frac{\text{d} M}{\text{d} t} = \mathcal{K} \frac{M_{\rm shell}}{t_{\rm inf}},
\end{equation}
where $M_{\rm shell}$ is the mass of an infalling shell with width $\Delta R$, $t_{\rm inf}$ is the time for the shell to accrete onto $R_{200c}$, and $\mathcal{K}$ is a scaling factor. 
In an infinitesimal shell ($\Delta R = {\rm d}r$) located at $R_{v_{min}}$, the shell mass is $M_{\rm shell}= 4\pi \rho(R_{v_{min}}) R_{v_{min}}^2 {\rm d}r$. If $t_{\rm inf}={\rm d}r/{\rm v}_{min}$ and $\mathcal{K}$ = 1, Eqs. (\ref{eq:mar}) and (\ref{eq:martheo}) are equivalent. Below, we derive $\mathcal{K}$ for thick shells of finite width.

Deriving the MAR from Eq. (\ref{eq:mar}) depends on estimates of the mass in the infalling shell and the infall time. The caustic method provides  the mass of the infalling shell from observations \citep{Diaferio1997,Diaferio99,Serra2011,Pizzardo23}. To calibrate this technique, we  compute the true mass of the shell by summing the mass of dark matter, gas, stars, and black holes in a shell of width $\Delta R$. 

There is no direct observational route to the infall time because $v_{inf}$ is currently inaccessible to observation.  Although $t_{\rm inf}$ depends on redshift, it is fairly independent of cluster mass (see Fig. \ref{fig:tinf} below). Thus, the combination of $M_{\rm shell}$ from observations and $t_{\rm inf}$ derived from simulations yield a MAR based on cluster observations.

To compute the infall time, we solve the equation of radial infall with a nonconstant acceleration derived from the true gravitational potential of the cluster. The parameters of the equation are the initial infall velocity of the infalling shell, $v_{\rm inf}$, the center of the infalling shell equivalent to the radial location of $v_{\rm inf}$, $R_{v_{min}}$, and the radius that defines the virialized region of clusters, $R_{200c}$ (Sect. \ref{subsec:tinf}). 

We locate the infalling shell (Sect. \ref{subsec:mshell}) from the simulated radial velocity profile of the cluster (Sect. \ref{subsec:vrad}). The boundaries of the infalling shell are  $R_{\rm shell,i}$ (inner cluster-centric  radius), and  $R_{\rm shell,o}$ (outer cluster-centric radius), where the smoothed radial velocity is a fraction $A$ of $v_{min}$. 

To link MARs in Eq. \ref{eq:mar} with MAR$_{\rm t}$, we compute MARs with a range of $A$ = 0.45--0.95. Smaller $A$ overlaps the virialized regions of clusters. Larger values of $A$
approach the ideal, theoretical limit in Eq. (\ref{eq:martheo}). However, MAR estimates  for these thin shells are less robust because the number of simulated particles in the infalling shell is small.

We compare MAR$^{C}$ and MAR$^{3D}$, the MARs estimated according to Eq. (\ref{eq:mar}) respectively using $M_{\rm shell}^C$ and $M_{\rm shell}^{3D}$, the caustic and 3D shell masses, for $\mathcal{K}$ = 1. Averaging over redshift, the median ratio between MAR$^{3D}$ and MAR$^{C}$, $r_{\rm 3c}$, increases monotonically from $r_{3c} = 0.92\pm 0.05$ ($A=0.45$) to $r_{3c} = 1.09\pm 0.04$ ($A=0.95$). The error in $r_{3c}$ is the interquartile range of the 11 ratios at the redshifts we consider. We adopt  $A=0.72$, where $r_{\rm 3c} = 1.00\pm 0.04$, as the best value. The choice of $A = 0.72$ provides a large enough galaxy sample for application of the caustic technique. The shaded blue vertical regions in Fig. \ref{fig:vrad} show shells with $A= 0.72$ for three redshifts: $z=0.01$, $z=0.62$, and $z=1.04$ (left to right, respectively). 

\begin{figure}
    \centering
    \includegraphics[width=\columnwidth]{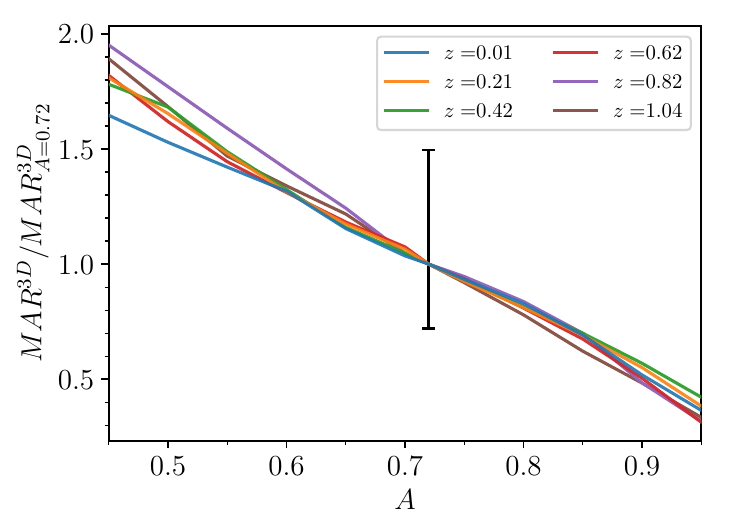}
    \caption{Relation between MAR and MAR$_t$. Curves show ${\rm MAR}^{3D}(A)/{\rm MAR}^{3D}(A=0.72)$ as a function of $A$ in the range $0.45-0.95$ for six redshifts from $z=0.01$ to $z=1.04$. The error bar shows the median interquartile range of the ratio at $A=0.72$.}
    \label{fig:fracv}
\end{figure}

Finally, we connect MARs estimated in extended shells to MAR$_{\rm t}$ and estimate the appropriate value for $\mathcal{K}$. Colored lines in Fig. \ref{fig:fracv} show the ratio ${\rm MAR}^{3D}(A) / {\rm MAR}^{3D}(A=0.72)$ as a function of $A$ at six different redshifts from $z=0.01$ to $z=1.04$ for $\mathcal{K} = 1$. The error bar shows the interquartile range of the ratios at $A=0.72$.  The estimated MARs decrease as $A$ increases.
  
When $A=0.95$, the radial velocity is nearly constant  across the shell and is (0.97--0.98)$v_{min}$. This thin shell provides an approximation to MAR$_{\rm t}$.
Averaging over the 11 redshifts, the ratio ${\rm MAR}^{3D}(A=0.95)/{\rm MAR}^{3D}(A = 0.72)$ is $\mathcal{K}=0.35$. This value provides a plausible scaling between MAR$_t$ and the $M_{\rm shell}$s and MARs estimated from cluster observations with $A=0.72$.

\section{$M_{\rm shell}$ and $t_{\rm inf}$ from 3D and caustic mass profiles}
\label{sec:mshelltinf}

Computation of the MAR of clusters requires an estimate of the mass of the infalling shell, $M_{\rm shell}$, and of the infall time for the cluster to accrete the shell, $t_{\rm inf}$. Section \ref{subsec:mshell} describes the determination of  $M_{\rm shell}$. Section \ref{subsec:tinf} describes the derivation of the infall time, $t_{\rm inf}$.

\subsection{The infalling shell and its mass $M_{\rm shell}$}
\label{subsec:mshell}

The average radial velocity profiles (Sect. \ref{subsec:vrad}; Fig. \ref{fig:vrad}) are the basis for determining the cluster-centric radius of the  infalling shell. 
The radial velocity profiles in Sect. \ref{subsec:vrad} show a clear infall pattern at  radii $\sim (2-3)R_{200c}$ where the minimum radial velocity occurs. Table \ref{table:rng_inf} lists the average minimum radial velocity, $v_{min}$, and its radial location, $R_{v_{min}}$, at each redshift. 

Based on the discussion of Section \ref{sec:recipe} we identify the inner  ($R_{\rm shell,i}$) and outer ($R_{\rm shell,o}$) boundaries of the infalling shell where the smoothed radial velocity is $0.72 v_{min}$ ($A=0.72$; shaded blue vertical regions in Fig. \ref{fig:vrad}). Table \ref{table:rng_inf} (fifth column) shows the radial location of the infalling shell at each redshift. The cluster-centric distance of the shells decreases as redshift increases: at low redshifts, $z\lesssim 0.4$, the shells are located in the range $\sim (1.6-3.4)R_{200c}$; at high redshift, $z \gtrsim 0.73$, the shells are in the range $\sim (1.3-2.6)R_{200c}$. This variation results from the decrease of $R_{v_{min}}$ with increasing  redshift (Sec. \ref{subsec:vrad}, see Fig. \ref{fig:vminvsrmin}). The width of the infalling shell, $\sim (1.2-1.3)R_{200c}$, is nearly redshift independent.

For each simulated cluster, we compute the true mass, $M_{\rm shell}^{3D}(A=0.72)$, and caustic observable mass, $M_{\rm shell}^{C}(A=0.72)$, of the infalling shell.  For the appropriate radial range (Table \ref{table:rng_inf}) we compute two true and caustic masses that measure $M(<R_{\rm shell,i};A=0.72)$, the total mass within the inner boundary of the shell, and $M(<R_{\rm shell,o};A=0.72)$, the total mass within the outer boundary of the  shell. The observable mass of the infalling shell is then the difference in these masses $M_{\rm shell}(A=0.72)=M(<R_{\rm shell,o};A=0.72)-M(<R_{\rm shell,i};A=0.72)$.

\begin{figure}
    \centering
    \includegraphics[width=\columnwidth]{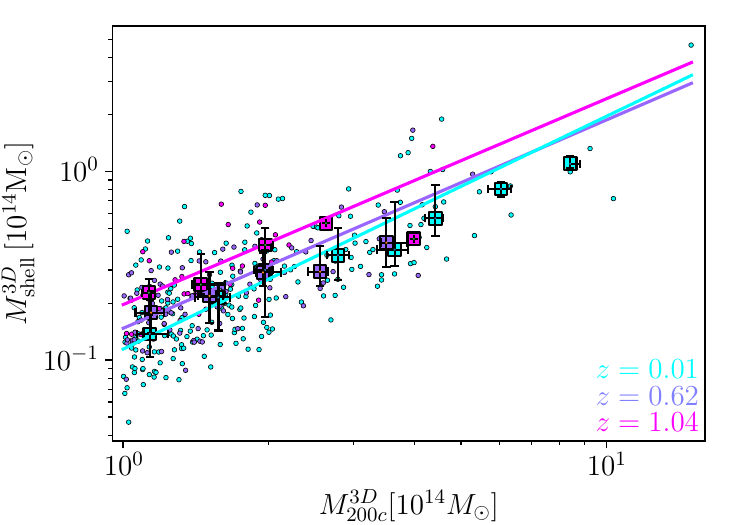}
    \caption{ $M_{\rm shell}^{3D}$ as a function of $M_{200c}^{3D}$. Points show $M_{\rm shell}^{3D}$ as a function of $M_{200c}^{3D}$ in the three redshift bins $z=0.01$, 0.62, and 1.04 (cyan, violet, and magenta, respectively). Squares with error bars indicate the median and interquartile range in eight fixed logarithmic mass bins. Lines show power law fits to the data.}
    \label{fig:mshellvsmass}
\end{figure}

Fig. \ref{fig:mshellvsmass} shows the true $M_{\rm shell}^{3D}=\mathcal{K} M_{\rm shell}^{3D}(A=0.72)$ as a function of the true $M_{200c}^{3D}$  at three different redshifts: $z=0.01$, $z=0.62$, and $z=1.04$ (cyan, violet, and magenta, respectively). The squares with error bars show the medians and interquartile ranges of the simulated data in eight logarithmic mass bins covering  the range $(1-12.6)\cdot 10^{14}$M$_\odot$. The lines show a power law fit to the data.

$M_{\rm shell}^{3D}$ and $M_{200c}^{3D}$ are strongly correlated at every redshift: a change of one order of magnitude in mass leads to a comparable change in the mass of the infalling shell. Kendall's test gives a correlation index of $\sim 0.44$ with associated p-values  in the range $\sim 10^{-5}-10^{-28}$. This correlation is expected in the hierarchical cluster formation model, because at fixed redshift higher mass clusters reside within greater overdensities and thus there is more mass in their accreting shells.
Conversely, $M_{\rm shell}$ is not strongly correlated with redshift. Although the fits show that $M_{\rm shell}^{3D}$  generally increases by $\sim 20-60\%$ from $z=0.01$ to $z=1.04$, the error bars show that this correlation is not statistically significant. 

\begin{figure}
    \centering
    \includegraphics[width=\columnwidth]{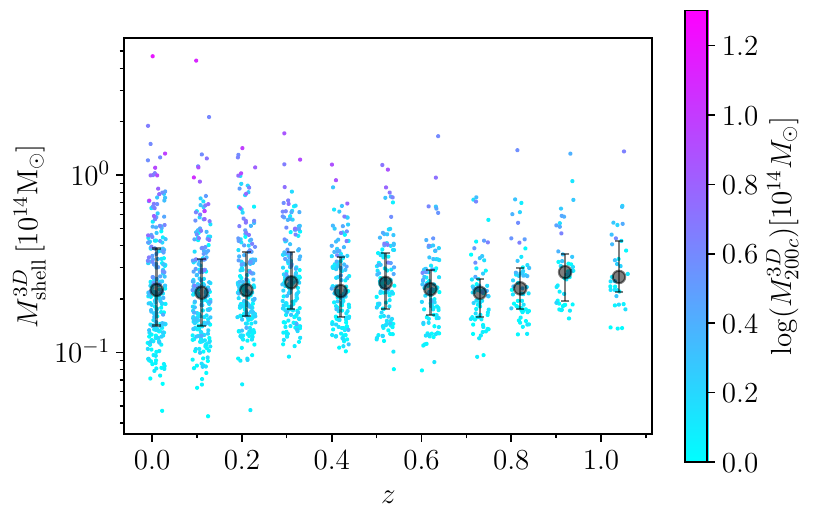}
    \caption{$M_{\rm shell}^{3D}$ as a function of redshift. Points show $M_{\rm shell}^{3D}$ of individual clusters as a function of redshift, colour coded by $M_{200c}^{3D}$. The points include clusters over the entire redshift range. We introduce an artificial jitter on the exact redshifts to make the dependence of $M_{\rm shell}^{3D}$ on $M_{200c}^{3D}$ more evident. Black circles with error bars show the median and interquartile range of $M_{\rm shell}^{3D}$ in each redshift bin.}
    \label{fig:mshellvsrmin}
\end{figure}

Coloured points in Fig. \ref{fig:mshellvsrmin} show $M_{\rm shell}^{3D}$ as a function of redshift for individual clusters in 11 redshift bins. Data are coded by cluster mass $M_{200c}^{3D}$. The 11 black circles with error bars indicate the median and  interquartile range of $M_{\rm shell}^{3D}$ in each redshift bin. 

Figure \ref{fig:mshellvsrmin} shows the same trend observed in Fig. \ref{fig:mshellvsmass} between $M_{\rm shell}^{3D}$ and $M_{200c}^{3D}$. At fixed redshift $M_{\rm shell}^{3D}$  increases with $M_{200c}^{3D}$: a change of $\sim 1$ order of magnitude in $M_{200c}^{3D}$ corresponds to an analogous change in $M_{\rm shell}^{3D}$. The median $M_{\rm shell}^{3D}$s, denoted by the black circles, are uncorrelated with redshift. 

The  dependence of $M_{\rm shell}$ and redshift is expected in the standard model of structure formation and evolution. In Appendix \ref{app} we explain how a complex interplay between the density and mass profiles as a function of redshift along with  the slightly different cluster  mass distribution at various redshifts tends to suppress any correlation between $M_{\rm shell}$ and redshift.

\begin{figure}
    \centering
    \includegraphics[width=\columnwidth]{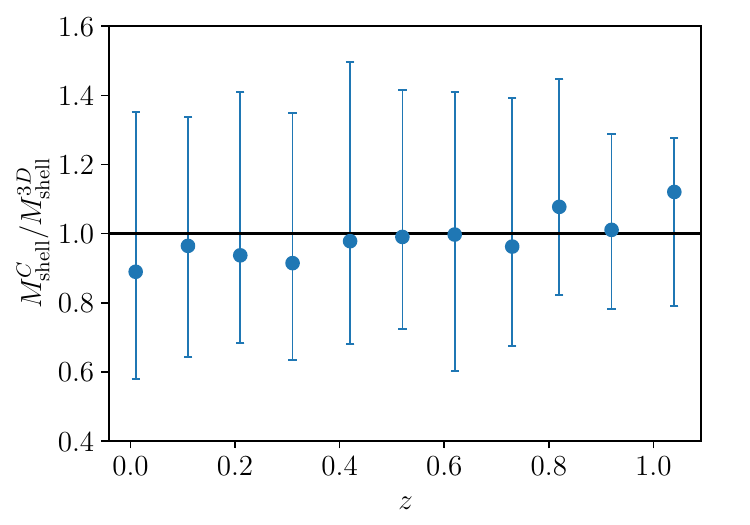}
    \caption{Median and interquartile ranges of the ratio between the caustic and  true $M_{\rm shell}$ as a function of redshift.}
    \label{fig:msh_cau3d}
\end{figure}

The caustic technique provides robust estimates of the mass profile of clusters in the infalling shell \citep{Pizzardo23}.  Points with error bars in Fig. \ref{fig:msh_cau3d} show the median and the interquartile range of individual  ratios between the caustic and true mass $M_{\rm shell}^{C} /M_{\rm shell}^{3D}$ as a function of redshift. Analogously to $M_{\rm shell}^{3D}$, we define $M_{\rm shell}^{C}=\mathcal{K} M_{\rm shell}^{C}(A=0.72)$.
Because the infalling region is closer to the cluster center as the redshift increases (Table \ref{table:rng_inf}, fifth column), the median ratio slowly rises from 0.9 at $z = 0$ to 1.1 at $z = 1.04$.\footnote{\citep{Pizzardo23} show that on average in the radial range $(0.6-4.2)R_{200c}$ the caustic mass is within the 10\% of the true mass. However, the caustic to true mass ratio slowly decreases from the lower to the upper end of the calibrated range.} The typical dispersion in the ratio is $\sim$ 34\%. 
Thus the caustic technique slightly overestimates or underestimates the mass at smaller and larger cluster-centric distances, respectively. The trend is not statistically significant:  Kendall's $\tau$ correlation coefficient is $\lesssim 0.05$. The caustic mass $M_{\rm shell}^{C}$ is an unbiased estimate of $M_{\rm shell}^{3D}$. Furthermore $M_{\rm shell}^{C}$ has the same dependence on mass, redshift, and $R_{v_{min}}$ as $M_{\rm shell}^{3D}$ on average.

\subsection{The infall time $t_{\rm inf}$}
\label{subsec:tinf}

Based on the full 3D cluster properties at each redshift, we compute $t_{\rm inf}$, the time for an infalling shell to accrete onto the virialized region of the cluster. We derive $t_{\rm inf}$ by tracking the radial infall with nonconstant acceleration driven by the cluster gravitational potential.
The locations of the minima in the average radial velocity profiles (Sect. \ref{subsec:vrad})  set the limit of the radial range where we compute the infall time. We ultimately employ the infall times based only on the  true full 3D data because the uncertainty in these measures derived by application of the caustic method are too large. 

We compute the time for the  center of the infalling shell  to reach   $R_{200c}$ as a measure of $t_{\rm inf}$. We measure the infall time starting from the radius of the minimum infall velocity, $R_{v_{min}}$ (Figure \ref{fig:vrad} and Table \ref{table:rng_inf}).  In other words, $t_{\rm inf}$ is the time required for the center of the infalling shell initially located at $R_{v_{min}}$ to reach $R_{200c}$.

Over the radial range  $R_{v_{min}}$ to $R_{200c}$, the gravitational acceleration  changes substantially. To compute $t_{\rm inf}$ of each cluster in each redshift bin we use an iterative procedure.

We divide the radial range $(R_{200c}, R_{v_{min}})$ into $N+1=101$ bins. The radial step between two contiguous bins is $\Delta r = (R_{v_{min}}-R_{200c})/N$. The $N+1$ steps $n=0,1,...,N$ correspond to $r_0=R_{v_{min}}$, $r_1=R_{v_{min}}-\Delta r$, ..., $r_{N}=R_{200c}$.

At each step $n<N$, we calculate $\Delta t_n$, the time for the center of the shell to move from $r_n=R_{v_{min}}-n\Delta r $ to $r_{n+1}=R_{v_{min}}-(n+1)\Delta r$. Within this small radial interval with $\Delta r\sim 0.01R_{200c}\approx 0.01$Mpc the gravitational acceleration is nearly constant. From this acceleration and the initial infall velocity of the shell, $a_n$ and $v_n$ respectively, we compute $\Delta t_n$ as the positive (physical) solution of the equation
\begin{equation}\label{eq:dt}
\frac{a_n}{2} \Delta t_n^2 + v_n \Delta t_n + \Delta r = 0,
\end{equation}
that is\footnote{In Eq. (\ref{eq:dt}), $a_n$ and $v_n$ are negative. The alternative mathematical solution to that in Eq. (\ref{eq:soldt}) returns a positive numerator, because $\sqrt{v_n^2-2a_n\Delta r} > -v_n$, and hence a negative (unphysical) $\Delta t_n$. }
\begin{equation}\label{eq:soldt}
    \Delta t_n = \frac{-v_n - \sqrt{v_n^2-2a_n\Delta r}}{a_n}.
\end{equation}

We obtain the acceleration of the infalling shell by computing the cluster gravitational potential $\phi(r)$ based on the true cluster shell density profile $\rho(r)$. The Poisson equation for an isolated spherical system with shell density profile $\rho_I(r)$ is:
\begin{equation}\label{eq:acc}
    \phi(r) = -4\pi G \left[\frac{1}{r} \int_{0}^r \rho_I(\Tilde{r})\Tilde{r}^2\, \text{d}\Tilde{r} +\int_{r}^{+\infty} \rho_I(\Tilde{r})\Tilde{r}\, \text{d}\Tilde{r}  \right].
\end{equation}
The uniform cosmological background density, $\braket{\rho(z)}=\Omega_M(z)\rho_c(z)$, exerts no net gravitational effect. Thus we compute the gravitational potential  based on  the mass density fluctuations  by replacing $\rho_I(r)$ with $\rho(r)-\braket{\rho}$. The second integral is finite;  at sufficiently large cluster-centric distances, $\sim 10R_{200c}$ (see Appendix \ref{app}), the correlated cluster density is $\sim 0$. We replace the upper limit of the second integral of Eq. (\ref{eq:acc}) with $10R_{200c}$.

From the true cluster potential $\phi(r)$ we compute the gravitational acceleration induced by the cluster, $a(r)$, by simple differentiation: $a(r)=-\frac{\text{d}}{\text{d} r}\phi(r)$. At each step $n$, we set $a_n$ (Eq. (\ref{eq:dt}) to the value of $a(r)$ at the position of the center of the infalling shell, $a_n = a(R_{v_{min}}-n\Delta r)$. 

At the first step $n=0$, the initial velocity is  $v_0=v_{\rm inf}$, where $v_{\rm inf}$ is the average cluster radial velocity of the infalling shell at that redshift (fourth column of Table \ref{table:rng_inf}). 
At each  succeeding step $n$ we increment the initial infall velocity of the shell by taking a constant acceleration in the small time step: $v_n=v_{n-1}+a_{n-1}\Delta t_{n-1}$. 

Application of Eq. (\ref{eq:soldt}) from $n=0$ to $n=N-1$ yields a set of $N$ time steps, $\Delta t_{n= \lbrace 0,...,N-1 \rbrace}$. 
The sum of these time steps is our estimate of the cluster infall time:
\begin{equation}\label{eq:tinf}
    t_{\rm inf} = \sum_{n=0}^{N-1} \Delta t_n.
\end{equation}

\begin{figure}
    \centering
    \includegraphics[width=\columnwidth]{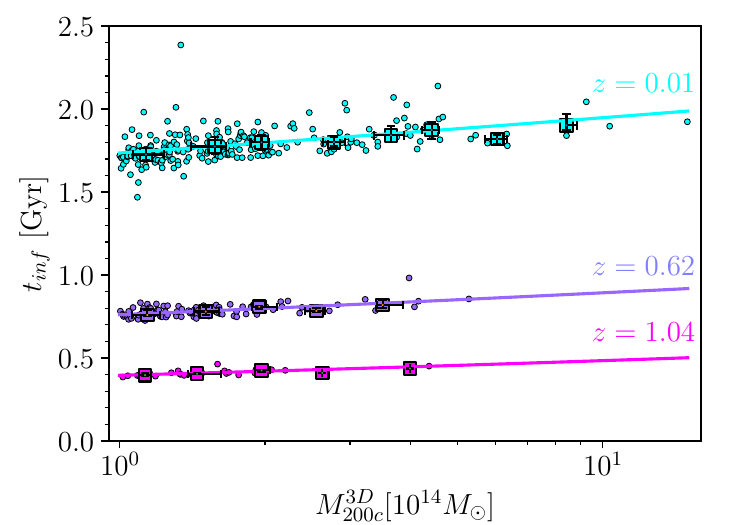}
    \caption{$t_{\rm inf}$ as a function of $M_{200c}^{3D}$. Points show $t_{\rm inf}$ of individual clusters as a function of their mass $M_{200c}^{3D}$ in the three redshift bins $z=0.01,0.62$, and 1.04 (cyan, violet, and magenta, respectively). Squares with error bars indicate the median and the interquartile range in eight fixed logarithmic bins of mass. Lines show power law fits to the data.}
    \label{fig:tinf}
\end{figure}

Figure \ref{fig:tinf} shows the resulting infall times for individual clusters as a function of $M_{200c}^{3D}$ in three redshift bins: $z=0.01,0.62,$ and 1.04  (cyan, violet, and magenta, respectively). The squares with error bars show the medians and interquartile ranges of the simulated data in eight logarithmic mass bins covering  the range $(1-12.6)\cdot 10^{14}$M$_\odot$ (as in Fig. \ref{fig:mshellvsmass}). The lines show a power law fit. 

The infall $t_{\rm inf}$ is  correlated with redshift. The increased cluster radial acceleration resulting from the larger cluster density at high redshift produces this correlation (Eq. \ref{eq:acc}). Equation \ref{eq:soldt} shows that the increased acceleration produces a corresponding  decrease in the infall time, $t_{\rm inf}$. 

The squares in Fig. \ref{fig:tinf} show the absence of correlation between $t_{\rm inf}$ and the cluster mass $M_{200c}^{3D}$ at each redshift.  The higher density of more massive clusters generates a larger acceleration decreasing $t_{\rm inf}$. However, more massive clusters are also more extended thus increasing the radial range $(R_{200c}, R_{v_{min}})$ and correspondingly increasing $t_{\rm inf}$. These effects result in a minimal dependence of $t_{\rm inf}$ on $M_{200c}^{3D}$.

\section{The Mass Accretion Rate}
\label{sec:results}

We apply Eq. \ref{eq:mar} to estimate the MARs at different redshifts. We compute MAR$^{3D}$, the true MAR based on $M_{\rm shell}^{3D}$ and $t_{\rm inf}$, and MAR$^{C}$, the caustic MAR based on $M_{\rm shell}^{C}$. As described earlier, we use 3D data to model the infall time. We compare MAR$^{C}$ with MAR$^{3D}$ to assess the caustic technique as a robust method for estimating  the true MAR.

\begin{figure}
    \centering
    \includegraphics[width=\columnwidth]{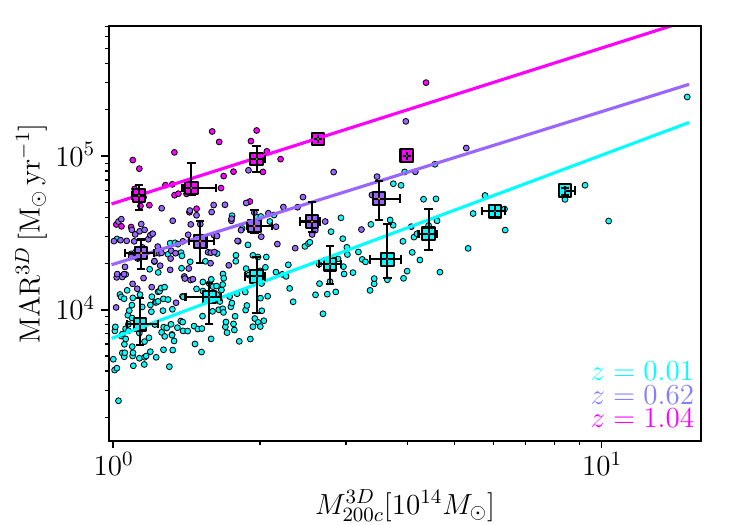}
    \caption{MAR$^{3D}$ as a function of $M_{200c}^{3D}$. Points show MAR$^{3D}$ as a function of $M_{200c}^{3D}$ in the three redshift bins $z=0.01, 0.62,$ and 1.04 (cyan, violet, and magenta, respectively). Squares with error bars show the median and interquartile range in eight fixed mass bins. Lines show power law fits to the data.}
    \label{fig:marvsmass}
\end{figure}

According to Eq. \ref{eq:mar}, the MAR$^{3D}$ of each individual cluster is the ratio between its respective $M_{\rm shell}^{3D}$ (see Sect. \ref{subsec:mshell}) and $t_{\rm inf}$ (see Sect. \ref{subsec:tinf}).  Fig. \ref{fig:marvsmass} shows the true MARs of individual clusters as a function of $M_{200c}^{3D}$ in three different redshift bins: $z=0.01, 0.62,$ and 1.04 (cyan, violet, and magenta, respectively). The squares with error bars show the medians and interquartile ranges of the simulated data in eight logarithmic mass bins covering  the range $(1-12.6)\cdot 10^{14}$M$_\odot$. The lines show a power law fit. 

The MARs increase both with increasing mass and with increasing redshift. 
A correlation between MAR and $M_{200c}$ is expected because more massive clusters tend to be surrounded by larger amounts of mass. Figure \ref{fig:marvsmass} shows that at fixed redshift a change of $\sim 1$ order of magnitude in $M_{200c}^{3D}$ corresponds to an analogous change in MAR$^{3D}$. This correlation with mass follows from the correlation between $M_{\rm shell}$ and $M_{200c}$ (Sect. \ref{subsec:mshell}): Figs. \ref{fig:marvsmass} and \ref{fig:mshellvsmass} show that the increase of MAR$^{3D}$ with $M_{200c}^{3D}$ is consistent with the increase of $M_{\rm shell}^{3D}$ with $M_{200c}^{3D}$. Fig. \ref{fig:tinf} shows that the infall time does not play a significant role in this correlation.

In the hierarchical clustering paradigm, clusters of fixed mass at higher redshift reside within denser regions and thus accrete faster than clusters of the same mass at lower redshift. Figure \ref{fig:marvsmass} shows that at fixed mass MAR$^{3D}$ increases by a factor $\sim 2.2$ from $z=0.01$ to $z=0.62$, and by a factor $\sim 5$ from $z=0.01$ to $z=1.04$. This effect originates from the anticorrelation between $t_{\rm inf}$ and redshift (Sect. \ref{subsec:tinf}): Figs. \ref{fig:marvsmass} and \ref{fig:tinf} show that the increase of MAR$^{3D}$ with redshift is consistent with this decrease of $t_{\rm inf}$ with redshift. 

We fit the individual MAR$^{3D}$s and $M_{200c}^{3D}$s at each redshift to the relation MAR$\,=a\left(M_{200c}^{3D}/10^{14}\text{M}_\odot \right)^b$ (Table \ref{table:fit}).
\begin{table}[htbp]
\begin{center}
\caption{\label{table:fit}  MAR$^{3D}$ as a function of $M_{200c}^{3D}$}
\begin{tabular}{ccc}
\hline
\hline
redshift & $a$ [$10^4$M$_\odot$yr$^{-1}$] & b \\
\hline
0.01 & $0.657\pm 0.003$ & $1.188\pm 0.002$ \\
0.11 & $0.766\pm 0.006$ & $1.145\pm 0.003$ \\
0.21 & $1.44\pm 0.01$ & $0.846\pm 0.003$ \\
0.31 & $1.434\pm 0.008$ & $0.929\pm 0.003$ \\
0.42 & $2.08\pm 0.01$ & $0.742\pm 0.003$ \\
0.52 & $2.24\pm 0.02$ & $0.805\pm 0.003$ \\
0.62 & $1.98\pm 0.03$ & $0.993\pm 0.007$\\
0.73 & $2.75\pm 0.05$ & $0.60\pm 0.01$\\
0.82 & $3.5\pm 0.2$ & $0.87\pm 0.02$\\
0.92 & $4.1\pm 0.3$ & $0.73\pm 0.03$\\
1.04 & $4.9\pm 0.4$ & $1.01\pm 0.02$\\
\hline
 \end{tabular}
 \end{center}
\end{table}  
The fits  show the expected tight correlation between MAR and redshift. The coefficient $a$ that measures the MAR at fixed mass increases with redshift by a factor $\sim 7$ from $z=0.01$ to $z=1.04$. The slope of the power law is essentially redshift independent. The average of the 11 slopes $b$'s yields a mean slope $\bar{b}=0.90\pm 0.18$ in the redshift range $0.01-1.04$.

We fit the analytic relation proposed by \citet{Fakhouri2010} to the MAR as a function of redshift:
\begin{equation}\label{eq:fak}
    \text{MAR}^{3D} = U [\text{M}_\odot \text{yr}^{-1}] \left(\frac{M_{200c}^{3D}}{10^{12}\text{M}_\odot} \right)^V (1+Wz)\sqrt{\Omega_{m0}(1+z)^3+\Omega_{\Lambda 0}}.
\end{equation}
Because of the limited mass range we sample at greater redshifts, we fix $V=\bar{b}$ and we fit the median MAR$^{3D}$ as a function of the median $M_{200c}^{3D}$ and the redshift. The resulting coefficients are: $U=128\pm 13$, $V=\bar{b}=0.90\pm 0.18$, and $W=1.69\pm 0.34$.  We compare these results with earlier work in Sect. \ref{sec:discussion}.

\begin{figure}
    \centering
    \includegraphics[width=\columnwidth]{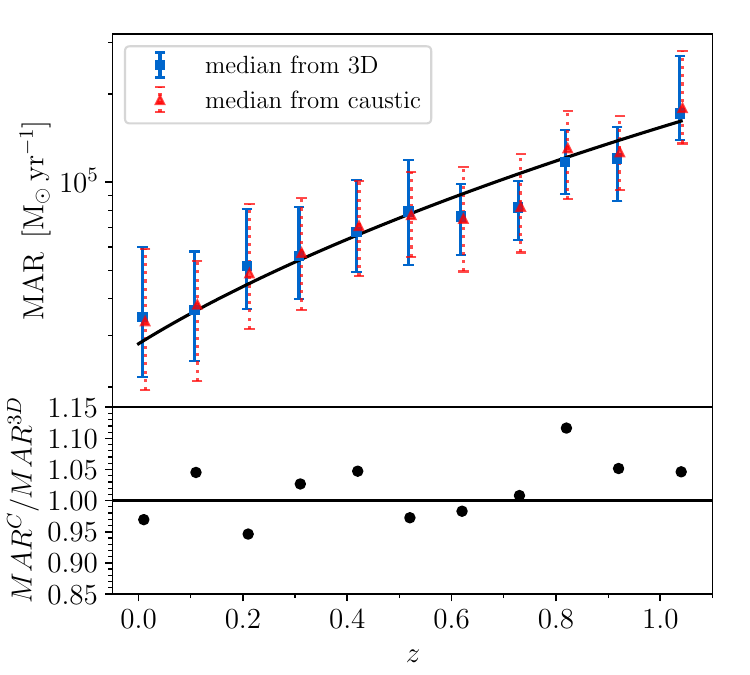}
    \caption{Comparison between MAR$^C$ and MAR$^{3D}$. Upper panel: Median true (blue squares) and caustic (red triangles) MARs as a function of redshift. The solid blue and dotted red error bars show the interquartile ranges of the true and caustic MARs, respectively. The black curve shows a fit to Eq. (\ref{eq:fak}). Lower panel: Points show the median ratio between the caustic and true MARs as a function of redshift.}
    \label{fig:mar3dcau}
\end{figure}

The MAR$^C$ of individual clusters is the ratio between $M_{\rm shell}^{C}$ and the infall time $t_{\rm inf}^{\rm fit}(M_{200c}^C)$ (Sect. \ref{subsec:tinf}), where we base the computation on the cluster caustic mass $M_{200c}^C$. We compute the cluster infall time $t_{\rm inf}^{\rm fit}(M_{200c}^C)$ by evaluating the power law fit of the individual infall times at the cluster caustic mass $M_{200c}^C$ as a function of mass derived by true profiles at the given redshift (Fig. \ref{fig:tinf}). The upper panel of Fig. \ref{fig:mar3dcau} shows the median and interquartile range of the true (blue squares and solid error bars) and caustic (red triangles and dotted error bars) MARs as a function of redshift. The black curve is a fit of Eq. (\ref{eq:fak}) to the MARs as a function of redshift. We fix the mass to the median mass of the cluster sample. The residuals around the fit are, on average, only $\sim -0.6\%$. Points in the lower panel show the ratios between the median values of the caustic and true MARs.

The figure shows that MAR$^C$ is a robust platform for estimating the true MAR$^{3D}$ at every redshift. The median MAR$^C$s are within  15\% of the median MAR$^{3D}$s at each redshift. The caustic MARs are also unbiased relative to the true MARs. Thus the caustic technique provides accurate and robust estimates of the MARs of clusters in the  redshift range $0.01-1.04$.

\section{Discussion}
\label{sec:discussion}

We use the Illustris TNG300-1 simulation to estimate the MAR of galaxy clusters based on the radial velocity profile of cluster galaxies and on the cluster total mass profile. The caustic technique provides robust and unbiased estimates of the true MARs  derived from 3D data over the  redshift range $0.01-1.04$.

We next (Sect. \ref{subsec:previous}) compare the Illustris results with previous work on simulated and observed clusters.  
In Sect. \ref{subsec:mardm} we assess the bias between the galaxy MARs and MARs derived from the dark matter halos for the same sample of clusters drawn from the IllustrisTNG simulations. In Sect. \ref{subsec:future} we discuss future simulations and observational applications of the MAR.

\subsection{Comparison with Previous Results}\label{subsec:previous}

The dynamically motivated MAR recipe we develop  differs significantly from previous  merger tree approaches \citep[e.g.,][]{mcbride2009,Fakhouri2010,vandenbosch14,Correa15b,Diemer2017sparta2,Diemer18}. In contrast with a merger tree procedure that is not directly observable, the approach we outline allows  the estimation of the MAR of real clusters and comparison with the true MARs of  comparable simulated systems. 

Most previous theoretical investigations of the MAR are based on N-body dark matter only simulations. These studies employ merger trees that trace the mass accretion of a halo at $z=0$ back in time. The merger trees follow  the change in mass between the halo descendant on the main branch identified at $z_i>0$ and its most massive progenitor for $z_i+\Delta z$ where $\Delta z$ is the time step.
The details of the simulation including the halo fragmentation algorithm and the choice of mass and time step may affect the results \citep{Diemer18,Xhakaji19}. Some studies \citep[e.g.,][]{vandenbosch14,Correa15a,Correa15b} are based on the analytic or semi-analytic extended Press-Schechter formalism \citep{Bond1991} calibrated with N-body simulations.

\begin{figure}
    \centering
    \includegraphics[width=\columnwidth]{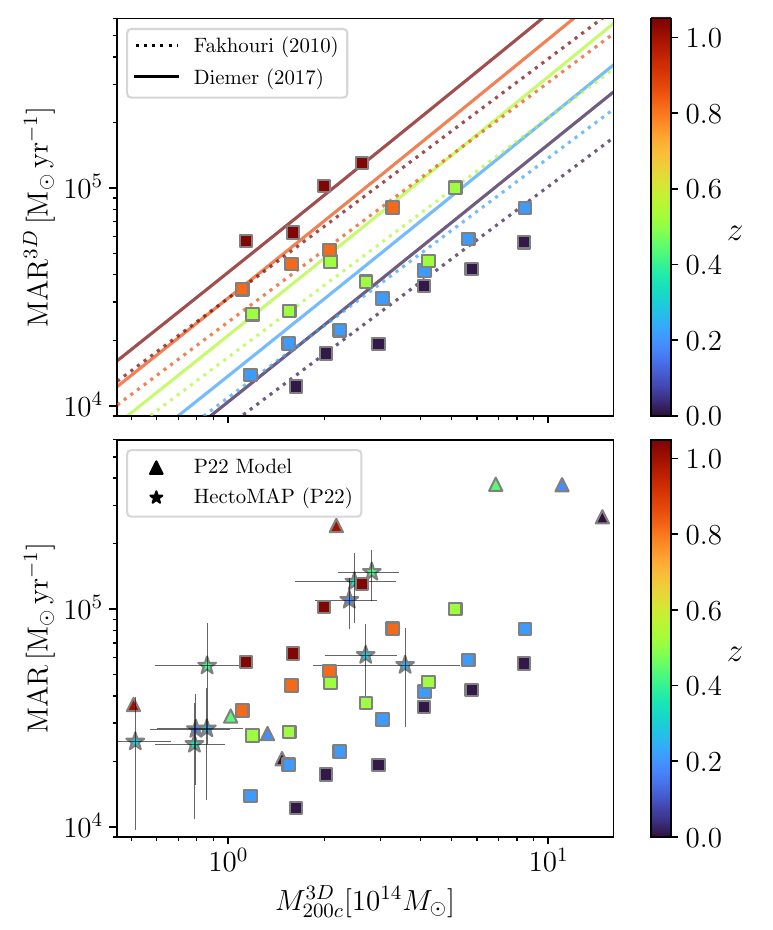}
    \caption{Comparison between our MARs and previously published results. Upper panel: Squares show MAR$^{3D}$s from Fig. \ref{fig:marvsmass} as a function of $M_{200c}^{3D}$ and colour coded by redshift from dark blue to dark red as the redshift increases from bin to bin:, $z=0.01,0.21,0.52,0.82,$ and $z=1.04$. Lines show MARs from merger trees (\citet{Fakhouri2010} (dotted line)  and \citet{Diemer2017sparta2} (solid line). 
    Lower panel: Squares again  show the true MARs from TNG300-1 as a function of $M_{200c}^{3D}$. Triangles show the MAR$^{3D}$s from \citet{Pizzardo2022} as a function of $M_{200c}^{3D}$ in four redshift bins: $z=0.01,0.19,0.44$, and $z=1.00$.  Stars show the MARs of ten observed stacked clusters from the HectoMAP survey \citep{Pizzardo2022}. }
    \label{fig:marcompa}
\end{figure}
The upper panel of Fig. \ref{fig:marcompa} compares MARs from Fig. \ref{fig:marvsmass} (squares) with two merger tree calculations based on
N-body simulations by \citet[][Eq. 2]{Fakhouri2010}\footnote{We show the numerical fit of \citet{Fakhouri2010} to their MARs.} (dotted lines) and \citet[][Eqs. 9-10]{Diemer2017sparta2} (solid lines).

\citet{Fakhouri2010} extend the  \citet{mcbride2009} investigation  of the Millennium simulation \citep{springel2005simulations} to  Millennium II \citep{boylank09}. For the mass of each FoF halo, they take the sum of the masses of their subhalos. They then compute the MARs from one snapshot to the next. The results by \citet{Fakhouri2010} are in excellent agreement with the earlier results of \citet{mcbride2009} and the later results of \citet[][see their Fig. 12]{vandenbosch14} and of \citet{Correa15a,Correa15b}. 

\citet{Diemer2017sparta2} use a large set of $\Lambda$CDM N-body simulations run with GADGET2 \citep{springel2005simulations}. They use  $M_{200m}$ as  a halo mass proxy and compute the MARs at time steps comparable with the halo dynamical time \citep{Diemer17i}\footnote{In Fig. \ref{fig:marcompa} we account for the differing Hubble parameters \citet{Fakhouri2010} and TNG300-1. We display the  \citet{Diemer2017sparta2}  results using the Colossus toolkit \citep{Diemer18} with the  TNG300-1 cosmology.}. 

Merger-tree MARs depend on the subhalo finder, the merger tree builder, and the definition of MAR. The difference between the two merger tree models in the upper panel of Fig. \ref{fig:marcompa} shows the impact of these underlying differences. At each time step a merger tree MAR  generally results from the difference between the mass of the descendant and the mass of the most massive progenitor. The most massive progenitor may  not be the main branch progenitor \citep[see][for more details]{Fakhouri2010} and thus the  MAR is a lower limit. The halo mass definition may also vary. In the \citet{Fakhouri2010} model the mass is the sum of the masses of the subhalos; in \citet{Diemer2017sparta2} the mass definition is $M_{200m}$, the mass enclosed within a sphere centered on the halo center with matter density equal to 200 times the background matter density. The choice of time step can also  affect the MARs \citep{Xhakaji19}. In Sect. \ref{subsec:mardm}, we demonstrate that the use of dark matter only simulations may also bias the MARs toward lower values, but the bias is small.

\begin{figure}
    \centering
    \includegraphics[width=\columnwidth]{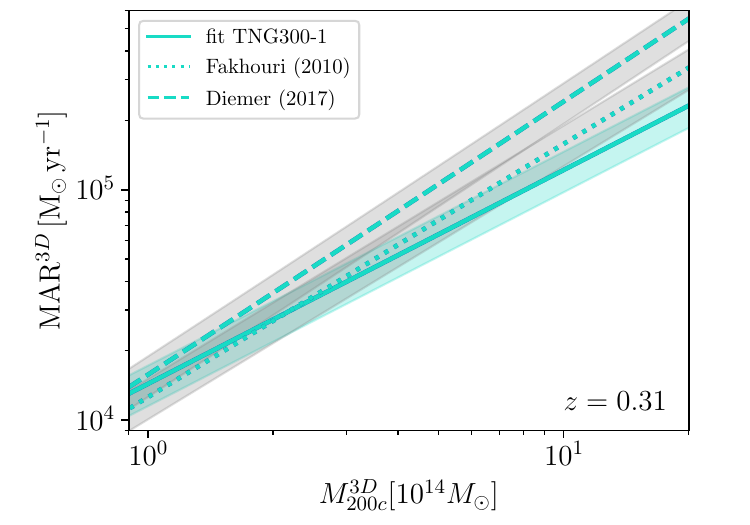}
    \caption{Comparison between fits to our MARs and previously published fits. The solid, dotted, and dashed lines show the fit of the analytic relation proposed by \citet{Fakhouri2010} to the Illustris MARs (Eq. (\ref{eq:fak}), see Sect. \ref{sec:results}), \citet{Fakhouri2010}, and \citet{Diemer2017sparta2}, respectively, for  $z=0.31$. The shadowed turquoise band shows the typical  $1\sigma$ IllustrisTNG scatter. The gray shadowing indicates a similar assumed scatter for the merger tree models (grey bands). }
    \label{fig:marcompafit}
\end{figure}

Fig. \ref{fig:marcompafit} compares the fits to the Illustris MARs using Eq. (\ref{eq:fak}) (see Sect. \ref{sec:results}) (solid line) with \citet{Fakhouri2010} (dotted line) and \citet{Diemer2017sparta2} (dashed line), at $z=0.31$. 
In all models, the MAR increases with cluster mass. Because \citet{Fakhouri2010} and \citet{Diemer2017sparta2} do not report errors in their fits we assume  fractional errors comparable with ours (shadowed areas). The slope of the IllustrisTNG MAR  then agrees with \citet{Fakhouri2010} and \citet{Diemer2014}. When averaged over the entire redshift range, the slope of the Illustris MARs as a function of $M_{200c}^{3D}$  is $0.90\pm 0.18$ (Sect. \ref{sec:results}) in agreement with the slopes of 1.1-1.2 obtained by \citet{Fakhouri2010} and \citet{Diemer2017sparta2}.

The IllustrisTNG MARs are consistent with merger tree MARs at every redshift (upper panel of Fig. \ref{fig:marcompa}). Assuming similar fractional errors for \citet{Fakhouri2010} and \citet{Diemer2017sparta2} and IllustrisTNG, the 0--50\% difference in the rates is usually within the 1$\sigma$ error and always within the 2$\sigma$ error. The shaded regions in Fig. \ref{fig:marcompafit} indicate the general consistency of the results. The difference between the IllustrisTNG and merger tree MARs increases as the redshift increases, but so does the error (Table \ref{table:fit}). The parameter $W$ of Eq. (\ref{eq:fak}) characterizes the redshift-dependence of the MAR. The IllustrisTNG simulations yield  $W=1.69 \pm 0.34$ (Sect. \ref{sec:results}); \citet{Fakhouri2010} obtain $W=1.17$. The difference is $\lesssim 2\sigma$. The qualitative agreement between the IllustrisTNG and merger tree approaches is reassuring given the substantial differences in the algorithms.

Based on the spherical accretion prescription proposed by \citet{deBoni2016}, \citet{pizzardo2020} develop the first systematic approach to estimating the MARs of real clusters.   \citet{pizzardo2020} compute the MAR as the ratio between the mass of an infalling shell and the infall time. Their approach is similar to the IllustrisTNG approach we follow, but they use a $\Lambda$CDM N-body dark matter only simulation.

\citet{Pizzardo2022} apply the  \citet{pizzardo2020} recipe to ten stacked clusters  from the HectoMAP redshift survey \citep{sohn2021hectomap,sohn2021cluster}. They also derive MARs based on the $\Lambda$CDM N-body simulation L-CoDECS \citep{Baldi2012CoDECS}. The bottom panel of Fig. \ref{fig:marcompa} shows the simulated MARs (triangles) and the observed  MARs of  HectoMAP stacked clusters (stars) as a function of mass and colour coded by redshift compared with  TNG300-1 (squares).

The HectoMAP MARs are consistent with the TNG300-1 MARs. The agreement between MARs of observed clusters and  TNG300-1 may reflect the use of simulated galaxies to calibrate the MAR derived from IllustrisTNG.

\subsection{The Dark matter MAR}\label{subsec:mardm}

Previous theoretical work on the MAR is based on N-body dark matter only simulations.
Galaxies are generally biased tracers of the underlying distribution of dark matter \citep[e.g.,][]{Kaiser84,Davis85,White87}. With Illustris TNG300-1 we can measure the bias by estimating the MAR for both galaxies and dark matter for the identical sample of clusters. We use 3D data from the simulation for this test. 

We begin by locating  the infalling shell based on the dark matter. We follow the procedure outlined in Sect. \ref{subsec:vrad} using the average radial velocity profiles of the dark matter particles. For each redshift bin, we compute the dark matter radial velocity profile of the individual clusters in 200 logarithmically spaced bins covering the radial range $(0-10)R_{200c}^{3D}$. We choose narrower binning for the dark matter profiles than we did for the galaxies because the number of dark matter particles is much larger than the number of galaxies. 

We compute a single mean radial velocity profile and  smooth it as we did  in Sect. \ref{subsec:vrad}. We identify the minimum radial velocity of the average profile, $v_{min}^{dm}$, and its cluster-centric location, $R_{v_{min}}^{dm}$. As in \ref{subsec:vrad}, the boundaries of the infalling are the cluster-centric distances where the average velocity is $0.75v_{min}^{dm}$.

The average radial velocity profiles of dark matter and galaxies agree at every redshift. The radial location of $v_{min}^{dm}$, $R_{v_{min}}^{dm}$, is on average $\sim 0.04\%$ ($\sim 7.9\%$ at most) smaller than $R_{v_{min}}$ based on the galaxies. The infall velocity based on the dark matter, $v_{\rm inf}^{dm}$, is on average $\sim 2.1\%$ ($\sim 5.9\%$ at most) less than $v_{\rm inf}$ determined from the galaxies.

We  compute the mass of the infalling shell, $M_{\rm shell}^{3D,dm}$, and the shell infall time, $t_{\rm inf}^{dm}$, from the dark matter field following Sects. \ref{subsec:mshell} and \ref{subsec:tinf} but based only on the mass profile of the dark matter component. To compute $M_{\rm shell}^{3D,dm}$ and $t_{\rm inf}^{dm}$ we multiply the dark matter profile by $(1+\Omega_{b0}/\Omega_{m0})$ to account for the baryonic fraction.  The resulting MAR$^{3D,dm}$ are then directly comparable with the MAR$^{3D}$ computed based on the total mass profile.

\begin{figure}
    \centering
    \includegraphics[width=\columnwidth]{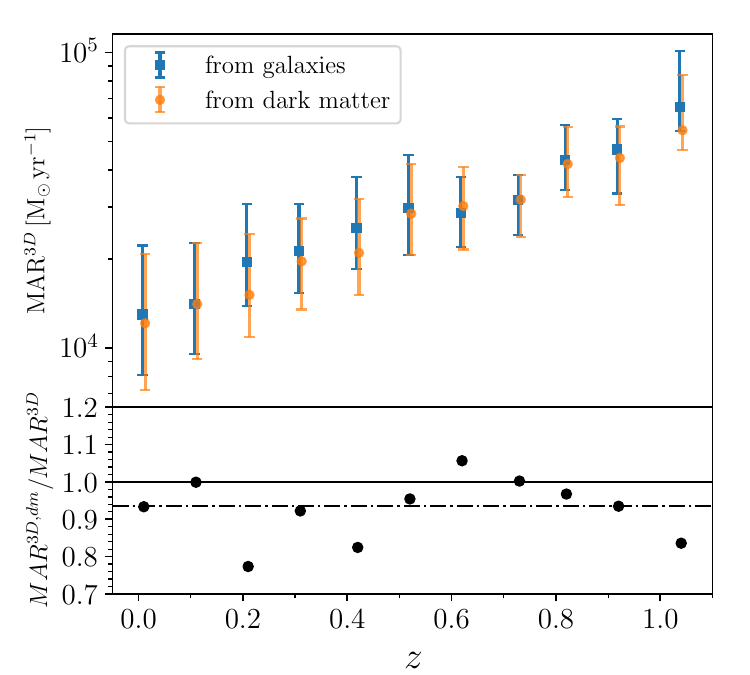}
    \caption{Comparison between MARs from galaxies and dark matter. Upper panel: Blue squares (orange dots) and blue (orange) error bars show the median and the interquartile range of the true MARs based on galaxies (dark matter) as a function  of redshift. Lower panel: Points show the median ratio between the dark matter and galaxy MARs as a function of redshift. The dash-dotted line shows the median ratio.}
    \label{fig:mar3ddm}
\end{figure}

Figure \ref{fig:mar3ddm} compares the dark matter and galaxy MARs. Blue squares and orange points in the upper panel show the median true MARs derived from galaxies and dark matter, respectively. The corresponding coloured error bars show the interquartile ranges of the MARs. Points in the lower panel show the median ratio between the dark matter and galaxies 3D MARs. The dash-dotted line show the global median.

On average, the dark matter MAR$^{3D,dm}$s are $\sim 6.5\%$ below the MARs based on the galaxies.
The scatter between the two MARs is $<30\%$, generally less than the uncertainty in the determination of the respective MARs. Thus galaxies are indeed biased tracers, but the bias is small on these scales. 

The MAR derived from galaxies should exceed the dark matter MAR because the clustering amplitude of galaxies relative to dark matter is larger at smaller scales and at higher redshift \citep{Davis83,Davis85,Jenkins98}. On the scale of the accretion region of galaxy clusters with redshift $z \lesssim1$, the galaxy clustering excess is, however, small \citep{Springel18}. Thus the bias between the galaxy and dark matter MARs derived from IllustrisTNG is also small.

\subsection{Future prospects}\label{subsec:future}

MARs of large samples of real and simulated clusters make the MARs a probe of cluster astrophysics \citep[e.g.,][]{Vitvitska_2002,gao2004sub,vandenbosch2005,kasun2005shapes,allgood2006shape,bett2007spin,ragone2010relation,Ludlow2013,Diemer2014,More2015,deBoni2016,Diemer2017sparta2} and cosmology \citep[e.g.,][]{diaferio2004outskirts,Mantz08,Mantz14,Mantz15,walker2019physics,cataneo2020tests}. The caustic technique \citep{Diaferio1997,Diaferio99,Serra2011} provides unbiased estimation of the true MAR in the wide redshift range $0.01-1.04$. Measurement of the MAR of real clusters  are currently limited to $z\lesssim 0.4$  \citep{pizzardo2020,Pizzardo2022}. 

Next generation wide-field spectroscopic surveys will observe the infall regions around large numbers of galaxy clusters with high sampling rates. These dense and deep spectroscopic surveys will provide the necessary observational baseline for measuring the MARs of thousands of clusters extending to higher redshifts. 

The  multi-object William Herschel Telescope Enhanced Area Velocity Explorer spectrograph on WHT \citep[WEAVE,][]{Balcells10,Dalton12,Dalton14,Dalton16} will explore  the infall regions of galaxy clusters and their connections to the cosmic web. The Weave Wide Field Cluster survey \citep[WWFC,][]{Cornwell22,Cornwell23} will measure thousands of galaxy spectra in and around 20 clusters with $0.04<z<0.07$ out to  radii $\lesssim 5 R_{200c}$. A deeper cluster survey will provide dense spectroscopic surveys of 100 clusters for redshift $\lesssim 0.5$, a new baseline for measuring the MAR in this redshift range.
Planned observations with the Prime Focus Spectrograph on Subaru \citep[PFS,][]{Tamura16} and the Maunakea Spectroscopic Explorer on CFHT \citep[MSE,][]{MSE19} will provide spectroscopic redshifts of hundreds to thousands of galaxy cluster members for thousands of individual clusters with $z\lesssim 0.6$.

The caustic technique \citep{Diaferio1997,Diaferio99} and weak gravitational lensing \citep{bartelmann2010gravitational,hoekstra2013masses,Umetsu20essay} will provide two independent measurements of cluster mass profiles extending to large radii. Neither the caustic technique nor weak lensing  relies on the assumption of dynamical equilibrium. These techniques  can thus be applied throughout the accretion region where dynamical equilibrium does not hold \citep{ludlow2009,bakels2020}. Present weak-lensing maps from HST and Subaru already provide cluster mass profiles up to $5.7$~Mpc for $\sim 20$ systems \citep{Umetsu11mp,Umetsu16}. Future facilities will extend these measurements to thousands of clusters. The VRO \citep{Ivezic19} and the Euclid mission \citep{Laureijs11,Sartoris16} will provide extended weak-lensing mass profiles for combination with extensive spectroscopy  samples of clusters with $z\lesssim 2$. 

The Illustris TNG300-1 MARs  are based on $ < 100$ clusters at $z> 0.52$. The mass distribution of TNG300-1 mostly samples clusters with $M_{200c}^{3D}\sim (1.2-2)\cdot 10^{14}$M$_\odot$. 
High redshift  massive clusters have larger MARs and place tight constraints on models of structure formation and evolution \citep[e.g.,][]{BBKS86,White02}. 
Larger volume hydrodynamical simulations, including MillenniumTNG \citep{MTNG22clus} with its 740~Mpc comoving size, will provide larger samples of the most massive systems up to higher redshift. The next generation of simulations should enable tracing of the MAR to redshifts $\gtrsim 1$. Extension of MAR determination to early epochs in cluster history will provide
new insights into the astrophysics of cluster formation and evolution.

\section{Conclusion}
\label{sec:conclusion}

We use the Illustris TNG300-1 simulation \citep{Pillepich18,Springel18,Nelson19} to compute the MAR of clusters of galaxies. The recipe, based on the dynamics of cluster galaxies,  computes the MAR as the ratio between the mass inside a spherical shell within the  cluster infall region and the time for the shell to reach the virialized region of the cluster. The method builds on the approach by \citet{deBoni2016} and \citet{pizzardo2020} and incorporates the caustic technique \citep{Diaferio1997,Diaferio99,Serra2011} that provides robust, unbiased estimates of the true MARs. IllustrisTNG refines estimation of clusters MARs. It permits derivation of  true MARs in a narrow shell, an approach that provides improved scaling between observational and true MARs. A major goal of the approach is direct application  to cluster observations. 

We use  1318 clusters extracted from TNG300-1  \citep{Pizzardo23}. This sample includes both the 3D and caustic mass profiles of each cluster. The clusters have median mass $M_{200c}^{3D}\sim (1.3-1.6)\cdot 10^{14}$M$_\odot$ and cover the redshift range $0.01-1.04$. We locate the infalling shell based on the average radial motion of cluster galaxies as a function of cluster-centric distance and redshift. We compute the infall time by solving the equation for radial infall of the infalling shell to $R_{200c}$ with nonconstant acceleration derived from the  true cluster  gravitational potential.

The MARs increase with increasing cluster mass and redshift. At fixed redshift, a change of $\sim 1$ order of magnitude in $M_{200c}$ yields a comparable increase in the MAR. This dependence tracks the increase of the mass of the infalling shell as a function of $M_{200c}$. At fixed mass, the MAR increases by a factor of $\sim 5$  from $z=0.01$ to $z=1.04$ because of the anticorrelation of the infall time with redshift. The correlations between the MAR and cluster mass and redshift are predicted by hierarchical structure formation scenarios.

The MARs from IllustrisTNG build on similar approaches based on N-body simulations \citep[e.g.,][]{pizzardo2020,Pizzardo2022}. In Illustris TNG300-1 we can test the dark matter MARs against the galaxy MARs for the identical set of simulated systems. The dark matter MARs are $\sim 6.5\%$ lower than the galaxy MARs reflecting the relative amplitudes of the clustering of galaxies and dark matter as a function of scale and redshift.

The IllustrisTNG MARs complement approaches based on merger trees which cannot be linked as directly to the observations \citep[e.g.,][]{mcbride2009,Fakhouri2010,Correa15b,vandenbosch14,Diemer2017sparta2,Diemer18}. 
The IllustrisTNG MARs lie within $2\sigma$ of the merger tree results. On average, the IllustrisTNG  MARs exceed the merger tree MARs by $\sim 50-70\%$; the difference increases with redshift. At fixed redshift, the dependence of the merger tree and Illustris MARs on cluster mass agree well. The IllustrisTNG MARs are remarkably consistent with available observations of the MAR as a function of redshift \citep{Pizzardo2022}.

IllustrisTNG enables  exploration of accretion by galaxy clusters with simulated galaxies. The approach provides a framework for obtaining robust MARs based on observed spectroscopic samples with $\gtrsim 200$ cluster members. Future spectroscopic (e.g. with instruments like WEAVE, PFS, and MSE) and weak lensing (e.g. with facilities like Euclid and VRO) surveys will provide these samples. The next generation of large volume hydrodynamical simulations including MillenniumTNG will guide the interpretation of observations of the MAR for higher redshift and and an extended cluster mass range. Determination of  MARs from the combined large datasets and enhanced simulations will test   models of formation and evolution of cosmic structure.

\begin{acknowledgements}
We thank Jubee Sohn for insightful discussions. 
Insightful comments from an anonymous reviewer helped us to refine our approach considerably and to provide more accurate MAR estimates.
M.P. and I.D. acknowledge the support of the Canada Research Chair Program and the Natural Sciences and Engineering Research Council of Canada (NSERC, funding reference number RGPIN-2018-05425).
The Smithsonian Institution supports the research of M.J.G. and S.J.K. A.D. acknowledges partial support from the grant InDark of the Italian National Institute of Nuclear Physics (INFN).
Part of the analysis was performed with the computer resources of INFN in Torino and of the University of Torino.
This research has made use of NASA's Astrophysics Data System Bibliographic Services.\\
All of the primary TNG simulations have been run on the Cray XC40 Hazel Hen supercomputer at the High Performance Computing Center Stuttgart (HLRS) in Germany. They have been made possible by the Gauss Centre for Supercomputing (GCS) large-scale project proposals GCS-ILLU and GCS-DWAR. GCS is the alliance of the three national supercomputing centres HLRS (Universitaet Stuttgart), JSC (Forschungszentrum Julich), and LRZ (Bayerische Akademie der Wissenschaften), funded by the German Federal Ministry of Education and Research (BMBF) and the German State Ministries for Research of Baden-Wuerttemberg (MWK), Bayern (StMWFK) and Nordrhein-Westfalen (MIWF). Further simulations were run on the Hydra and Draco supercomputers at the Max Planck Computing and Data Facility (MPCDF, formerly known as RZG) in Garching near Munich, in addition to the Magny system at HITS in Heidelberg. Additional computations were carried out on the Odyssey2 system supported by the FAS Division of Science, Research Computing Group at Harvard University, and the Stampede supercomputer at the Texas Advanced Computing Center through the XSEDE project AST140063.
\end{acknowledgements}

\bibliographystyle{aa}
\bibliography{main}

\begin{thebibliography}{95}
\expandafter\ifx\csname natexlab\endcsname\relax\def\natexlab#1{#1}\fi

\bibitem[{{Achitouv} {et~al.}(2014){Achitouv}, {Wagner}, {Weller}, \&
  {Rasera}}]{achitouv2014}
{Achitouv}, I., {Wagner}, C., {Weller}, J., \& {Rasera}, Y. 2014, \jcap, 2014,
  077

\bibitem[{{Adhikari} {et~al.}(2014){Adhikari}, {Dalal}, \&
  {Chamberlain}}]{adhikari2014}
{Adhikari}, S., {Dalal}, N., \& {Chamberlain}, R.~T. 2014, JCAP, 11, 019

\bibitem[{Allgood {et~al.}(2006)Allgood, Flores, Primack, Kravtsov, Wechsler,
  Faltenbacher, \& Bullock}]{allgood2006shape}
Allgood, B., Flores, R.~A., Primack, J.~R., {et~al.} 2006, \mnras, 367, 1781

\bibitem[{{Bakels} {et~al.}(2021){Bakels}, {Ludlow}, \& {Power}}]{bakels2020}
{Bakels}, L., {Ludlow}, A.~D., \& {Power}, C. 2021, \mnras, 501, 5948

\bibitem[{{Balcells} {et~al.}(2010){Balcells}, {Benn}, {Carter}, {Dalton},
  {Trager}, {Feltzing}, {Verheijen}, {Jarvis}, {Percival}, {Abrams}, {Agocs},
  {Brown}, {Cano}, {Evans}, {Helmi}, {Lewis}, {McLure}, {Peletier},
  {P{\'e}rez-Fournon}, {Sharples}, {Tosh}, {Trujillo}, {Walton}, \&
  {Westhall}}]{Balcells10}
{Balcells}, M., {Benn}, C.~R., {Carter}, D., {et~al.} 2010, in Society of
  Photo-Optical Instrumentation Engineers (SPIE) Conference Series, Vol. 7735,
  Ground-based and Airborne Instrumentation for Astronomy III, ed. I.~S.
  {McLean}, S.~K. {Ramsay}, \& H.~{Takami}, 77357G

\bibitem[{{Baldi}(2012)}]{Baldi2012CoDECS}
{Baldi}, M. 2012, \mnras, 422, 1028

\bibitem[{{Bardeen} {et~al.}(1986){Bardeen}, {Bond}, {Kaiser}, \&
  {Szalay}}]{BBKS86}
{Bardeen}, J.~M., {Bond}, J.~R., {Kaiser}, N., \& {Szalay}, A.~S. 1986, \apj,
  304, 15

\bibitem[{Bartelmann(2010)}]{bartelmann2010gravitational}
Bartelmann, M. 2010, Classical and Quantum Gravity, 27, 233001

\bibitem[{Bett {et~al.}(2007)Bett, Eke, Frenk, Jenkins, Helly, \&
  Navarro}]{bett2007spin}
Bett, P., Eke, V., Frenk, C.~S., {et~al.} 2007, \mnras, 376, 215

\bibitem[{{Bond} {et~al.}(1991){Bond}, {Cole}, {Efstathiou}, \&
  {Kaiser}}]{Bond1991}
{Bond}, J.~R., {Cole}, S., {Efstathiou}, G., \& {Kaiser}, N. 1991, \apj, 379,
  440

\bibitem[{Bower(1991)}]{bower1991}
Bower, R.~G. 1991, \mnras, 248, 332

\bibitem[{{Boylan-Kolchin} {et~al.}(2009){Boylan-Kolchin}, {Springel}, {White},
  {Jenkins}, \& {Lemson}}]{boylank09}
{Boylan-Kolchin}, M., {Springel}, V., {White}, S. D.~M., {Jenkins}, A., \&
  {Lemson}, G. 2009, \mnras, 398, 1150

\bibitem[{Cataneo \& Rapetti(2020)}]{cataneo2020tests}
Cataneo, M. \& Rapetti, D. 2020, in Modified Gravity: Progresses and Outlook of
  Theories, Numerical Techniques and Observational Tests (World Scientific),
  143--173

\bibitem[{Corasaniti \& Achitouv(2011)}]{corasaniti2011}
Corasaniti, P.~S. \& Achitouv, I. 2011, Phys. Rev. Lett., 106, 241302

\bibitem[{{Cornwell} {et~al.}(2023){Cornwell}, {Arag{\'o}n-Salamanca},
  {Kuchner}, {Gray}, {Pearce}, \& {Knebe}}]{Cornwell23}
{Cornwell}, D.~J., {Arag{\'o}n-Salamanca}, A., {Kuchner}, U., {et~al.} 2023,
  \mnras, 524, 2148

\bibitem[{{Cornwell} {et~al.}(2022){Cornwell}, {Kuchner},
  {Arag{\'o}n-Salamanca}, {Gray}, {Pearce}, {Aguerri}, {Cui},
  {M{\'e}ndez-Abreu}, {Peralta de Arriba}, \& {Trager}}]{Cornwell22}
{Cornwell}, D.~J., {Kuchner}, U., {Arag{\'o}n-Salamanca}, A., {et~al.} 2022,
  \mnras, 517, 1678

\bibitem[{{Correa} {et~al.}(2015a){Correa}, {Wyithe}, {Schaye}, \&
  {Duffy}}]{Correa15a}
{Correa}, C.~A., {Wyithe}, J. S.~B., {Schaye}, J., \& {Duffy}, A.~R. 2015a,
  \mnras, 450, 1514

\bibitem[{{Correa} {et~al.}(2015b){Correa}, {Wyithe}, {Schaye}, \&
  {Duffy}}]{Correa15b}
{Correa}, C.~A., {Wyithe}, J. S.~B., {Schaye}, J., \& {Duffy}, A.~R. 2015b,
  \mnras, 450, 1521

\bibitem[{{Dalton}(2016)}]{Dalton16}
{Dalton}, G. 2016, in Astronomical Society of the Pacific Conference Series,
  Vol. 507, Multi-Object Spectroscopy in the Next Decade: Big Questions, Large
  Surveys, and Wide Fields, ed. I.~{Skillen}, M.~{Balcells}, \& S.~{Trager}, 97

\bibitem[{{Dalton} {et~al.}(2014){Dalton}, {Trager}, {Abrams}, {Bonifacio},
  {L{\'o}pez Aguerri}, {Middleton}, {Benn}, {Dee}, {Say{\`e}de}, {Lewis},
  {Pragt}, {Pico}, {Walton}, {Rey}, {Allende Prieto}, {Pe{\~n}ate}, {Lhome},
  {Ag{\'o}cs}, {Alonso}, {Terrett}, {Brock}, {Gilbert}, {Ridings}, {Guinouard},
  {Verheijen}, {Tosh}, {Rogers}, {Steele}, {Stuik}, {Tromp}, {Jasko}, {Kragt},
  {Lesman}, {Mottram}, {Bates}, {Gribbin}, {Fernando Rodriguez}, {Delgado},
  {Martin}, {Cano}, {Navarro}, {Irwin}, {Lewis}, {Gonzalez Solares},
  {O'Mahony}, {Bianco}, {Zurita}, {ter Horst}, {Molinari}, {Lodi}, {Guerra},
  {Vallenari}, \& {Baruffolo}}]{Dalton14}
{Dalton}, G., {Trager}, S., {Abrams}, D.~C., {et~al.} 2014, in Society of
  Photo-Optical Instrumentation Engineers (SPIE) Conference Series, Vol. 9147,
  Ground-based and Airborne Instrumentation for Astronomy V, ed. S.~K.
  {Ramsay}, I.~S. {McLean}, \& H.~{Takami}, 91470L

\bibitem[{Dalton {et~al.}(2012)Dalton, Trager, Abrams, Carter, Bonifacio,
  Aguerri, MacIntosh, Evans, Lewis, Navarro, {et~al.}}]{Dalton12}
Dalton, G., Trager, S.~C., Abrams, D.~C., {et~al.} 2012, in Ground-based and
  Airborne Instrumentation for Astronomy IV, Vol. 8446, SPIE, 220--231

\bibitem[{{Davis} {et~al.}(1985){Davis}, {Efstathiou}, {Frenk}, \&
  {White}}]{Davis85}
{Davis}, M., {Efstathiou}, G., {Frenk}, C.~S., \& {White}, S.~D.~M. 1985, \apj,
  292, 371

\bibitem[{{Davis} \& {Peebles}(1983)}]{Davis83}
{Davis}, M. \& {Peebles}, P.~J.~E. 1983, \apj, 267, 465

\bibitem[{{De Boni} {et~al.}(2016){De Boni}, {Serra}, {Diaferio}, {Giocoli}, \&
  {Baldi}}]{deBoni2016}
{De Boni}, C., {Serra}, A.~L., {Diaferio}, A., {Giocoli}, C., \& {Baldi}, M.
  2016, \apj, 818, 188

\bibitem[{De~Simone {et~al.}(2011)De~Simone, Maggiore, \&
  Riotto}]{desimone2011}
De~Simone, A., Maggiore, M., \& Riotto, A. 2011, \mnras, 418, 2403

\bibitem[{{Diaferio}(1999)}]{Diaferio99}
{Diaferio}, A. 1999, \mnras, 309, 610

\bibitem[{Diaferio(2004)}]{diaferio2004outskirts}
Diaferio, A. 2004, Outskirts of Galaxy Clusters (IAU C195): Intense Life in the
  Suburbs No. 195 (Cambridge University Press)

\bibitem[{{Diaferio} \& {Geller}(1997)}]{Diaferio1997}
{Diaferio}, A. \& {Geller}, M.~J. 1997, \apj, 481, 633

\bibitem[{{Diemer}(2017a)}]{Diemer17i}
{Diemer}, B. 2017a, \apjs, 231, 5

\bibitem[{{Diemer}(2018)}]{Diemer18}
{Diemer}, B. 2018, \apjs, 239, 35

\bibitem[{{Diemer} \& {Kravtsov}(2014)}]{Diemer2014}
{Diemer}, B. \& {Kravtsov}, A.~V. 2014, \apj, 789, 1

\bibitem[{{Diemer} {et~al.}(2017b){Diemer}, {Mansfield}, {Kravtsov}, \&
  {More}}]{Diemer2017sparta2}
{Diemer}, B., {Mansfield}, P., {Kravtsov}, A.~V., \& {More}, S. 2017b, \apj,
  843, 140

\bibitem[{{Fakhouri} {et~al.}(2010){Fakhouri}, {Ma}, \&
  {Boylan-Kolchin}}]{Fakhouri2010}
{Fakhouri}, O., {Ma}, C.-P., \& {Boylan-Kolchin}, M. 2010, \mnras, 406, 2267

\bibitem[{Gao {et~al.}(2004)Gao, White, Jenkins, Stoehr, \&
  Springel}]{gao2004sub}
Gao, L., White, S. D.~M., Jenkins, A., Stoehr, F., \& Springel, V. 2004,
  \mnras, 355, 819

\bibitem[{Giocoli {et~al.}(2012)Giocoli, Tormen, \& Sheth}]{giocoli2012}
Giocoli, C., Tormen, G., \& Sheth, R.~K. 2012, \mnras, 422, 185

\bibitem[{{Gunn} \& {Gott}(1972)}]{Gunn1972}
{Gunn}, J.~E. \& {Gott}, J.~Richard, I. 1972, \apj, 176, 1

\bibitem[{Hoekstra {et~al.}(2013)Hoekstra, Bartelmann, Dahle, Israel, Limousin,
  \& Meneghetti}]{hoekstra2013masses}
Hoekstra, H., Bartelmann, M., Dahle, H., {et~al.} 2013, \ssr, 177, 75

\bibitem[{{Ivezi{\'c}} {et~al.}(2019){Ivezi{\'c}}, {Kahn}, {Tyson}, {Abel},
  {Acosta}, {Allsman}, {Alonso}, {AlSayyad}, {Anderson}, {Andrew}, {Angel},
  {Angeli}, {Ansari}, {Antilogus}, {Araujo}, {Armstrong}, {Arndt}, {Astier},
  {Aubourg}, {Auza}, {Axelrod}, {Bard}, {Barr}, {Barrau}, {Bartlett}, {Bauer},
  {Bauman}, {Baumont}, {Bechtol}, {Bechtol}, {Becker}, {Becla}, {Beldica},
  {Bellavia}, {Bianco}, {Biswas}, {Blanc}, {Blazek}, {Blandford}, {Bloom},
  {Bogart}, {Bond}, {Booth}, {Borgland}, {Borne}, {Bosch}, {Boutigny},
  {Brackett}, {Bradshaw}, {Brandt}, {Brown}, {Bullock}, {Burchat}, {Burke},
  {Cagnoli}, {Calabrese}, {Callahan}, {Callen}, {Carlin}, {Carlson},
  {Chandrasekharan}, {Charles-Emerson}, {Chesley}, {Cheu}, {Chiang}, {Chiang},
  {Chirino}, {Chow}, {Ciardi}, {Claver}, {Cohen-Tanugi}, {Cockrum}, {Coles},
  {Connolly}, {Cook}, {Cooray}, {Covey}, {Cribbs}, {Cui}, {Cutri}, {Daly},
  {Daniel}, {Daruich}, {Daubard}, {Daues}, {Dawson}, {Delgado}, {Dellapenna},
  {de Peyster}, {de Val-Borro}, {Digel}, {Doherty}, {Dubois},
  {Dubois-Felsmann}, {Durech}, {Economou}, {Eifler}, {Eracleous}, {Emmons},
  {Fausti Neto}, {Ferguson}, {Figueroa}, {Fisher-Levine}, {Focke}, {Foss},
  {Frank}, {Freemon}, {Gangler}, {Gawiser}, {Geary}, {Gee}, {Geha}, {Gessner},
  {Gibson}, {Gilmore}, {Glanzman}, {Glick}, {Goldina}, {Goldstein}, {Goodenow},
  {Graham}, {Gressler}, {Gris}, {Guy}, {Guyonnet}, {Haller}, {Harris},
  {Hascall}, {Haupt}, {Hernandez}, {Herrmann}, {Hileman}, {Hoblitt}, {Hodgson},
  {Hogan}, {Howard}, {Huang}, {Huffer}, {Ingraham}, {Innes}, {Jacoby}, {Jain},
  {Jammes}, {Jee}, {Jenness}, {Jernigan}, {Jevremovi{\'c}}, {Johns}, {Johnson},
  {Johnson}, {Jones}, {Juramy-Gilles}, {Juri{\'c}}, {Kalirai}, {Kallivayalil},
  {Kalmbach}, {Kantor}, {Karst}, {Kasliwal}, {Kelly}, {Kessler}, {Kinnison},
  {Kirkby}, {Knox}, {Kotov}, {Krabbendam}, {Krughoff}, {Kub{\'a}nek},
  {Kuczewski}, {Kulkarni}, {Ku}, {Kurita}, {Lage}, {Lambert}, {Lange},
  {Langton}, {Le Guillou}, {Levine}, {Liang}, {Lim}, {Lintott}, {Long},
  {Lopez}, {Lotz}, {Lupton}, {Lust}, {MacArthur}, {Mahabal}, {Mandelbaum},
  {Markiewicz}, {Marsh}, {Marshall}, {Marshall}, {May}, {McKercher}, {McQueen},
  {Meyers}, {Migliore}, {Miller}, {Mills}, {Miraval}, {Moeyens}, {Moolekamp},
  {Monet}, {Moniez}, {Monkewitz}, {Montgomery}, {Morrison}, {Mueller},
  {Muller}, {Mu{\~n}oz Arancibia}, {Neill}, {Newbry}, {Nief}, {Nomerotski},
  {Nordby}, {O'Connor}, {Oliver}, {Olivier}, {Olsen}, {O'Mullane}, {Ortiz},
  {Osier}, {Owen}, {Pain}, {Palecek}, {Parejko}, {Parsons}, {Pease},
  {Peterson}, {Peterson}, {Petravick}, {Libby Petrick}, {Petry},
  {Pierfederici}, {Pietrowicz}, {Pike}, {Pinto}, {Plante}, {Plate}, {Plutchak},
  {Price}, {Prouza}, {Radeka}, {Rajagopal}, {Rasmussen}, {Regnault}, {Reil},
  {Reiss}, {Reuter}, {Ridgway}, {Riot}, {Ritz}, {Robinson}, {Roby}, {Roodman},
  {Rosing}, {Roucelle}, {Rumore}, {Russo}, {Saha}, {Sassolas}, {Schalk},
  {Schellart}, {Schindler}, {Schmidt}, {Schneider}, {Schneider}, {Schoening},
  {Schumacher}, {Schwamb}, {Sebag}, {Selvy}, {Sembroski}, {Seppala}, {Serio},
  {Serrano}, {Shaw}, {Shipsey}, {Sick}, {Silvestri}, {Slater}, {Smith},
  {Smith}, {Sobhani}, {Soldahl}, {Storrie-Lombardi}, {Stover}, {Strauss},
  {Street}, {Stubbs}, {Sullivan}, {Sweeney}, {Swinbank}, {Szalay}, {Takacs},
  {Tether}, {Thaler}, {Thayer}, {Thomas}, {Thornton}, {Thukral}, {Tice},
  {Trilling}, {Turri}, {Van Berg}, {Vanden Berk}, {Vetter}, {Virieux},
  {Vucina}, {Wahl}, {Walkowicz}, {Walsh}, {Walter}, {Wang}, {Wang}, {Warner},
  {Wiecha}, {Willman}, {Winters}, {Wittman}, {Wolff}, {Wood-Vasey}, {Wu},
  {Xin}, {Yoachim}, \& {Zhan}}]{Ivezic19}
{Ivezi{\'c}}, {\v{Z}}., {Kahn}, S.~M., {Tyson}, J.~A., {et~al.} 2019, \apj,
  873, 111

\bibitem[{{Jenkins} {et~al.}(1998){Jenkins}, {Frenk}, {Pearce}, {Thomas},
  {Colberg}, {White}, {Couchman}, {Peacock}, {Efstathiou}, \&
  {Nelson}}]{Jenkins98}
{Jenkins}, A., {Frenk}, C.~S., {Pearce}, F.~R., {et~al.} 1998, \apj, 499, 20

\bibitem[{{Kaiser}(1984)}]{Kaiser84}
{Kaiser}, N. 1984, \apjl, 284, L9

\bibitem[{Kasun \& Evrard(2005)}]{kasun2005shapes}
Kasun, S. \& Evrard, A.~E. 2005, \apj, 629, 781

\bibitem[{Lacey \& Cole(1993)}]{laceyCole93}
Lacey, C. \& Cole, S. 1993, \mnras, 262, 627

\bibitem[{Lau {et~al.}(2015)Lau, Nagai, Avestruz, Nelson, \&
  Vikhlinin}]{lau2015mass}
Lau, E.~T., Nagai, D., Avestruz, C., Nelson, K., \& Vikhlinin, A. 2015, \apj,
  806, 68

\bibitem[{{Laureijs} {et~al.}(2011){Laureijs}, {Amiaux}, {Arduini},
  {Augu{\`e}res}, {Brinchmann}, {Cole}, {Cropper}, {Dabin}, {Duvet}, {Ealet},
  {Garilli}, {Gondoin}, {Guzzo}, {Hoar}, {Hoekstra}, {Holmes}, {Kitching},
  {Maciaszek}, {Mellier}, {Pasian}, {Percival}, {Rhodes}, {Saavedra Criado},
  {Sauvage}, {Scaramella}, {Valenziano}, {Warren}, {Bender}, {Castander},
  {Cimatti}, {Le F{\`e}vre}, {Kurki-Suonio}, {Levi}, {Lilje}, {Meylan},
  {Nichol}, {Pedersen}, {Popa}, {Rebolo Lopez}, {Rix}, {Rottgering},
  {Zeilinger}, {Grupp}, {Hudelot}, {Massey}, {Meneghetti}, {Miller}, {Paltani},
  {Paulin-Henriksson}, {Pires}, {Saxton}, {Schrabback}, {Seidel}, {Walsh},
  {Aghanim}, {Amendola}, {Bartlett}, {Baccigalupi}, {Beaulieu}, {Benabed},
  {Cuby}, {Elbaz}, {Fosalba}, {Gavazzi}, {Helmi}, {Hook}, {Irwin}, {Kneib},
  {Kunz}, {Mannucci}, {Moscardini}, {Tao}, {Teyssier}, {Weller}, {Zamorani},
  {Zapatero Osorio}, {Boulade}, {Foumond}, {Di Giorgio}, {Guttridge}, {James},
  {Kemp}, {Martignac}, {Spencer}, {Walton}, {Bl{\"u}mchen}, {Bonoli},
  {Bortoletto}, {Cerna}, {Corcione}, {Fabron}, {Jahnke}, {Ligori}, {Madrid},
  {Martin}, {Morgante}, {Pamplona}, {Prieto}, {Riva}, {Toledo}, {Trifoglio},
  {Zerbi}, {Abdalla}, {Douspis}, {Grenet}, {Borgani}, {Bouwens}, {Courbin},
  {Delouis}, {Dubath}, {Fontana}, {Frailis}, {Grazian}, {Koppenh{\"o}fer},
  {Mansutti}, {Melchior}, {Mignoli}, {Mohr}, {Neissner}, {Noddle}, {Poncet},
  {Scodeggio}, {Serrano}, {Shane}, {Starck}, {Surace}, {Taylor},
  {Verdoes-Kleijn}, {Vuerli}, {Williams}, {Zacchei}, {Altieri}, {Escudero
  Sanz}, {Kohley}, {Oosterbroek}, {Astier}, {Bacon}, {Bardelli}, {Baugh},
  {Bellagamba}, {Benoist}, {Bianchi}, {Biviano}, {Branchini}, {Carbone},
  {Cardone}, {Clements}, {Colombi}, {Conselice}, {Cresci}, {Deacon}, {Dunlop},
  {Fedeli}, {Fontanot}, {Franzetti}, {Giocoli}, {Garcia-Bellido}, {Gow},
  {Heavens}, {Hewett}, {Heymans}, {Holland}, {Huang}, {Ilbert}, {Joachimi},
  {Jennins}, {Kerins}, {Kiessling}, {Kirk}, {Kotak}, {Krause}, {Lahav}, {van
  Leeuwen}, {Lesgourgues}, {Lombardi}, {Magliocchetti}, {Maguire}, {Majerotto},
  {Maoli}, {Marulli}, {Maurogordato}, {McCracken}, {McLure}, {Melchiorri},
  {Merson}, {Moresco}, {Nonino}, {Norberg}, {Peacock}, {Pello}, {Penny},
  {Pettorino}, {Di Porto}, {Pozzetti}, {Quercellini}, {Radovich}, {Rassat},
  {Roche}, {Ronayette}, {Rossetti}, {Sartoris}, {Schneider}, {Semboloni},
  {Serjeant}, {Simpson}, {Skordis}, {Smadja}, {Smartt}, {Spano}, {Spiro},
  {Sullivan}, {Tilquin}, {Trotta}, {Verde}, {Wang}, {Williger}, {Zhao},
  {Zoubian}, \& {Zucca}}]{Laureijs11}
{Laureijs}, R., {Amiaux}, J., {Arduini}, S., {et~al.} 2011, arXiv e-prints,
  arXiv:1110.3193

\bibitem[{{Ludlow}(2009)}]{ludlow2009}
{Ludlow}, A.~D. 2009, PhD thesis, University of Victoria, Canada

\bibitem[{{Ludlow} {et~al.}(2013){Ludlow}, {Navarro}, {Boylan-Kolchin}, {Bett},
  {Angulo}, {Li}, {White}, {Frenk}, \& {Springel}}]{Ludlow2013}
{Ludlow}, A.~D., {Navarro}, J.~F., {Boylan-Kolchin}, M., {et~al.} 2013, \mnras,
  432, 1103

\bibitem[{{Mantz} {et~al.}(2008){Mantz}, {Allen}, {Ebeling}, \&
  {Rapetti}}]{Mantz08}
{Mantz}, A., {Allen}, S.~W., {Ebeling}, H., \& {Rapetti}, D. 2008, \mnras, 387,
  1179

\bibitem[{{Mantz} {et~al.}(2014){Mantz}, {Allen}, {Morris}, {Rapetti},
  {Applegate}, {Kelly}, {von der Linden}, \& {Schmidt}}]{Mantz14}
{Mantz}, A.~B., {Allen}, S.~W., {Morris}, R.~G., {et~al.} 2014, \mnras, 440,
  2077

\bibitem[{{Mantz} {et~al.}(2015a){Mantz}, {von der Linden}, {Allen},
  {Applegate}, {Kelly}, {Morris}, {Rapetti}, {Schmidt}, {Adhikari}, {Allen},
  {Burchat}, {Burke}, {Cataneo}, {Donovan}, {Ebeling}, {Shandera}, \&
  {Wright}}]{Mantz15}
{Mantz}, A.~B., {von der Linden}, A., {Allen}, S.~W., {et~al.} 2015a, \mnras,
  446, 2205

\bibitem[{{Marshall} {et~al.}(2019){Marshall}, {Bolton}, {Bullock},
  {Burgasser}, {Chambers}, {DePoy}, {Dey}, {Flagey}, {Hill}, {Hillenbrand},
  {Huber}, {Li}, {Juneau}, {Kaplinghat}, {Mateo}, {McConnachie}, {Newman},
  {Petric}, {Schlegel}, {Sheinis}, {Shen}, {Simons}, {Strauss}, {Szeto},
  {Tran}, \& {Y{\`e}che}}]{MSE19}
{Marshall}, J., {Bolton}, A., {Bullock}, J., {et~al.} 2019, in \baas, Vol.~51,
  126

\bibitem[{{McBride} {et~al.}(2009){McBride}, {Fakhouri}, \& {Ma}}]{mcbride2009}
{McBride}, J., {Fakhouri}, O., \& {Ma}, C.-P. 2009, \mnras, 398, 1858

\bibitem[{{Meiksin}(1985)}]{Meiksin86}
{Meiksin}, A. 1985

\bibitem[{{More} {et~al.}(2015){More}, {Diemer}, \& {Kravtsov}}]{More2015}
{More}, S., {Diemer}, B., \& {Kravtsov}, A.~V. 2015, \apj, 810, 36

\bibitem[{{Musso} {et~al.}(2018){Musso}, {Cadiou}, {Pichon}, {Codis},
  {Kraljic}, \& {Dubois}}]{musso2018}
{Musso}, M., {Cadiou}, C., {Pichon}, C., {et~al.} 2018, \mnras, 476, 4877

\bibitem[{Nelson {et~al.}(2019)Nelson, Springel, Pillepich, Rodriguez-Gomez,
  Torrey, Genel, Vogelsberger, Pakmor, Marinacci, Weinberger,
  {et~al.}}]{Nelson19}
Nelson, D., Springel, V., Pillepich, A., {et~al.} 2019, Computational
  Astrophysics and Cosmology, 6, 1

\bibitem[{{O'Neil} {et~al.}(2021){O'Neil}, {Barnes}, {Vogelsberger}, \&
  {Diemer}}]{oneil21}
{O'Neil}, S., {Barnes}, D.~J., {Vogelsberger}, M., \& {Diemer}, B. 2021,
  \mnras, 504, 4649

\bibitem[{{Pakmor} {et~al.}(2023){Pakmor}, {Springel}, {Coles}, {Guillet},
  {Pfrommer}, {Bose}, {Barrera}, {Delgado}, {Ferlito}, {Frenk}, {Hadzhiyska},
  {Hern{\'a}ndez-Aguayo}, {Hernquist}, {Kannan}, \& {White}}]{MTNG22clus}
{Pakmor}, R., {Springel}, V., {Coles}, J.~P., {et~al.} 2023, \mnras, 524, 2539

\bibitem[{{Pillepich} {et~al.}(2018){Pillepich}, {Springel}, {Nelson}, {Genel},
  {Naiman}, {Pakmor}, {Hernquist}, {Torrey}, {Vogelsberger}, {Weinberger}, \&
  {Marinacci}}]{Pillepich18}
{Pillepich}, A., {Springel}, V., {Nelson}, D., {et~al.} 2018, \mnras, 473, 4077

\bibitem[{{Pizzardo} {et~al.}(2021){Pizzardo}, {Di Gioia}, {Diaferio}, {De
  Boni}, {Serra}, {Geller}, {Sohn}, {Rines}, \& {Baldi}}]{pizzardo2020}
{Pizzardo}, M., {Di Gioia}, S., {Diaferio}, A., {et~al.} 2021, \aap, 646, A105

\bibitem[{{Pizzardo} {et~al.}(2023){Pizzardo}, {Geller}, {Kenyon}, {Damjanov},
  \& {Diaferio}}]{Pizzardo23}
{Pizzardo}, M., {Geller}, M.~J., {Kenyon}, S.~J., {Damjanov}, I., \&
  {Diaferio}, A. 2023, A\&A, 675, A56

\bibitem[{{Pizzardo} {et~al.}(2022){Pizzardo}, {Sohn}, {Geller}, {Diaferio}, \&
  {Rines}}]{Pizzardo2022}
{Pizzardo}, M., {Sohn}, J., {Geller}, M.~J., {Diaferio}, A., \& {Rines}, K.
  2022, \apj, 927, 26

\bibitem[{Power {et~al.}(2011)Power, Knebe, \& Knollmann}]{power2011}
Power, C., Knebe, A., \& Knollmann, S.~R. 2011, \mnras, 419, 1576

\bibitem[{Press \& Schechter(1974)}]{press1974formation}
Press, W.~H. \& Schechter, P. 1974, \apj, 187, 425

\bibitem[{Ragone-Figueroa {et~al.}(2010)Ragone-Figueroa, Plionis, Merch{\'a}n,
  Gottl{\"o}ber, \& Yepes}]{ragone2010relation}
Ragone-Figueroa, C., Plionis, M., Merch{\'a}n, M., Gottl{\"o}ber, S., \& Yepes,
  G. 2010, \mnras, 407, 581

\bibitem[{Reiprich {et~al.}(2013)Reiprich, Basu, Ettori, Israel, Lovisari,
  Molendi, Pointecouteau, \& Roncarelli}]{reiprich2013outskirts}
Reiprich, T.~H., Basu, K., Ettori, S., {et~al.} 2013, \ssr, 177, 195

\bibitem[{{Rines} \& {Diaferio}(2006)}]{Rines2006CIRS}
{Rines}, K. \& {Diaferio}, A. 2006, \aj, 132, 1275

\bibitem[{{Rines} {et~al.}(2013){Rines}, {Geller}, {Diaferio}, \&
  {Kurtz}}]{Rines2013HeCS}
{Rines}, K., {Geller}, M.~J., {Diaferio}, A., \& {Kurtz}, M.~J. 2013, \apj,
  767, 15

\bibitem[{Rost {et~al.}(2021)Rost, Kuchner, Welker, Pearce, Stasyszyn, Gray,
  Cui, Dave, Knebe, Yepes, {et~al.}}]{rost2021threehundred}
Rost, A., Kuchner, U., Welker, C., {et~al.} 2021, \mnras, 502, 714

\bibitem[{{Sartoris} {et~al.}(2016){Sartoris}, {Biviano}, {Fedeli}, {Bartlett},
  {Borgani}, {Costanzi}, {Giocoli}, {Moscardini}, {Weller}, {Ascaso},
  {Bardelli}, {Maurogordato}, \& {Viana}}]{Sartoris16}
{Sartoris}, B., {Biviano}, A., {Fedeli}, C., {et~al.} 2016, \mnras, 459, 1764

\bibitem[{Savitzky \& Golay(1964)}]{Savitzky64}
Savitzky, A. \& Golay, M.~J. 1964, Analytical chemistry, 36, 1627

\bibitem[{{Schechter}(1980)}]{Schechter80}
{Schechter}, P.~L. 1980, \aj, 85, 801

\bibitem[{{Serra} {et~al.}(2011){Serra}, {Diaferio}, {Murante}, \&
  {Borgani}}]{Serra2011}
{Serra}, A.~L., {Diaferio}, A., {Murante}, G., \& {Borgani}, S. 2011, \mnras,
  412, 800

\bibitem[{Sheth \& Tormen(2002)}]{sheth2002}
Sheth, R.~K. \& Tormen, G. 2002, \mnras, 329, 61

\bibitem[{Silk(1974)}]{silk1974}
Silk, J. 1974, \apj, 193, 525

\bibitem[{{Sohn} {et~al.}(2021b){Sohn}, {Geller}, {Hwang}, {Diaferio}, {Rines},
  \& {Utsumi}}]{sohn2021cluster}
{Sohn}, J., {Geller}, M.~J., {Hwang}, H.~S., {et~al.} 2021b, \apj, 923, 143

\bibitem[{{Sohn} {et~al.}(2021a){Sohn}, {Geller}, {Hwang}, {Fabricant},
  {Moran}, \& {Utsumi}}]{sohn2021hectomap}
{Sohn}, J., {Geller}, M.~J., {Hwang}, H.~S., {et~al.} 2021a, \apj, 909, 129

\bibitem[{{Springel} {et~al.}(2018){Springel}, {Pakmor}, {Pillepich},
  {Weinberger}, {Nelson}, {Hernquist}, {Vogelsberger}, {Genel}, {Torrey},
  {Marinacci}, \& {Naiman}}]{Springel18}
{Springel}, V., {Pakmor}, R., {Pillepich}, A., {et~al.} 2018, \mnras, 475, 676

\bibitem[{Springel {et~al.}(2005)Springel, White, Jenkins, Frenk, Yoshida, Gao,
  Navarro, Thacker, Croton, Helly, {et~al.}}]{springel2005simulations}
Springel, V., White, S.~D., Jenkins, A., {et~al.} 2005, \nat, 435, 629

\bibitem[{Tamura {et~al.}(2016)Tamura, Takato, Shimono, Moritani, Yabe,
  Ishizuka, Ueda, Kamata, Aghazarian, Arnouts, {et~al.}}]{Tamura16}
Tamura, N., Takato, N., Shimono, A., {et~al.} 2016, in Ground-based and
  Airborne Instrumentation for Astronomy VI, Vol. 9908, International Society
  for Optics and Photonics, 99081M

\bibitem[{Tasitsiomi {et~al.}(2004)Tasitsiomi, Kravtsov, Gottlober, \&
  Klypin}]{Tasitsiomi_2004}
Tasitsiomi, A., Kravtsov, A.~V., Gottlober, S., \& Klypin, A.~A. 2004, \apj,
  607, 125

\bibitem[{{Umetsu}(2020)}]{Umetsu20essay}
{Umetsu}, K. 2020, \aapr, 28, 7

\bibitem[{{Umetsu} {et~al.}(2011){Umetsu}, {Broadhurst}, {Zitrin},
  {Medezinski}, {Coe}, \& {Postman}}]{Umetsu11mp}
{Umetsu}, K., {Broadhurst}, T., {Zitrin}, A., {et~al.} 2011, \apj, 738, 41

\bibitem[{{Umetsu} {et~al.}(2016){Umetsu}, {Zitrin}, {Gruen}, {Merten},
  {Donahue}, \& {Postman}}]{Umetsu16}
{Umetsu}, K., {Zitrin}, A., {Gruen}, D., {et~al.} 2016, \apj, 821, 116

\bibitem[{van~den Bosch(2002)}]{vandenbosch02}
van~den Bosch, F.~C. 2002, \mnras, 331, 98

\bibitem[{{van den Bosch} {et~al.}(2014){van den Bosch}, {Jiang}, {Hearin},
  {Campbell}, {Watson}, \& {Padmanabhan}}]{vandenbosch14}
{van den Bosch}, F.~C., {Jiang}, F., {Hearin}, A., {et~al.} 2014, \mnras, 445,
  1713

\bibitem[{Van Den~Bosch {et~al.}(2005)Van Den~Bosch, Tormen, \&
  Giocoli}]{vandenbosch2005}
Van Den~Bosch, F.~C., Tormen, G., \& Giocoli, C. 2005, \mnras, 359, 1029

\bibitem[{Vitvitska {et~al.}(2002)Vitvitska, Klypin, Kravtsov, Wechsler,
  Primack, \& Bullock}]{Vitvitska_2002}
Vitvitska, M., Klypin, A.~A., Kravtsov, A.~V., {et~al.} 2002, \apj, 581, 799

\bibitem[{Walker {et~al.}(2019)Walker, Simionescu, Nagai, Okabe, Eckert,
  Mroczkowski, Akamatsu, Ettori, \& Ghirardini}]{walker2019physics}
Walker, S., Simionescu, A., Nagai, D., {et~al.} 2019, \ssr, 215, 1

\bibitem[{Wechsler {et~al.}(2002)Wechsler, Bullock, Primack, Kravtsov, \&
  Dekel}]{Wechsler_2002}
Wechsler, R.~H., Bullock, J.~S., Primack, J.~R., Kravtsov, A.~V., \& Dekel, A.
  2002, \apj, 568, 52

\bibitem[{White(2002)}]{White02}
White, M. 2002, \apj Supplement Series, 143, 241

\bibitem[{{White} {et~al.}(1987){White}, {Davis}, {Efstathiou}, \&
  {Frenk}}]{White87}
{White}, S. D.~M., {Davis}, M., {Efstathiou}, G., \& {Frenk}, C.~S. 1987, \nat,
  330, 451

\bibitem[{White \& Rees(1978)}]{white1978}
White, S. D.~M. \& Rees, M.~J. 1978, \mnras, 183, 341

\bibitem[{{Xhakaj} {et~al.}(2019){Xhakaj}, {Leauthaud}, {Diemer}, \&
  {Behroozi}}]{Xhakaji19}
{Xhakaj}, E., {Leauthaud}, A., {Diemer}, B., \& {Behroozi}, P. 2019, Research
  Notes of the American Astronomical Society, 3, 169

\bibitem[{Zhang {et~al.}(2008)Zhang, Ma, \& Fakhouri}]{zhang2008}
Zhang, J., Ma, C.-P., \& Fakhouri, O. 2008, \mnras: Letters, 387, L13

\bibitem[{{Zhang} {et~al.}(2004){Zhang}, {Finoguenov}, {B{\"o}hringer},
  {Ikebe}, {Matsushita}, \& {Schuecker}}]{zhang2004}
{Zhang}, Y.~Y., {Finoguenov}, A., {B{\"o}hringer}, H., {et~al.} 2004, \aap,
  413, 49

\end{thebibliography}

\begin{appendix}
\section{The Redshift Dependence of  $M_{\rm shell}$}\label{app}

The fits in Fig. \ref{fig:mshellvsmass} (Sect. \ref{subsec:mshell}) indicate that $M_{\rm shell}^{3D}$ increases by $\sim 20-60\%$ from $z=0.01$ to $z=1.04$ depending on the cluster mass. Because of the large scatter, any correlation with redshift is statistically insignificant. 
Here we outline the reasons for this minimal dependence of $M_{\rm shell}$ on redshift. This result reflects an interplay between the large-scale  cluster density profiles  and the slightly different distributions of cluster masses sampled by the simulations  as a function of redshift.

\begin{figure}
    \centering
    \includegraphics[scale=0.85]{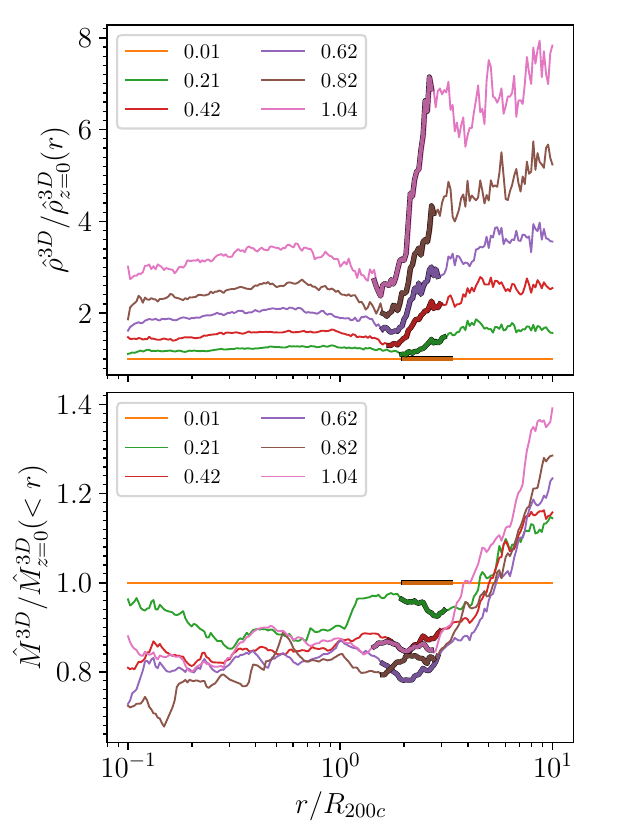}
    \caption{Upper panel: Ratio between the median shell density of clusters at the six redshifts $z=0.01,0.21,0.42,0.62,0.82,$ and 1.04 relative to $z=0.01$. Colour coding is shown in the legend. The thick section of each ratio shows the infalling shell (Sect. \ref{subsec:mshell}, Table \ref{table:rng_inf}). Bottom panel: Same as the upper panel, but for the median cumulative mass profile.}
    \label{fig:rhosh}
\end{figure}

The upper panel of Fig. \ref{fig:rhosh} shows the true median density profiles of clusters  ($\hat{\rho}^{3D}$)  relative to $z = 0.01$. The profiles are scaled to  $r/R_{200c}^{3D}$ in each redshift bin as noted in the legend. For each scaled profile the bold region indicates the range of radii of the infalling shell (fifth column of Table \ref{table:rng_inf}).

Clusters  at higher redshift are denser than  their lower redshift counterparts as expected.  In the inner equilibrium region, $\lesssim R_{200c}^{3D}$ (Sect. \ref{subsec:vrad}, Fig. \ref{fig:vrad}), the density ratios are roughly constant from one redshift bin to another. For $r \gtrsim R_{200c}$, the ratios reach  a minimum and then increase. At larger cluster-centric distances, the ratios reach the ratios of the average cosmological mass density at the relevant epochs. Subtracting the cosmological mean density from the profiles does not change the qualitative behavior of the relative profiles. The cluster density dominates the total  mass density for  $r \lesssim (6-7)R_{200c}$, a radius larger than the characteristic turnaround radius (Sect. \ref{subsec:vrad}). 

The infalling shells are outside the virialized region (thick section in each curve of Fig. \ref{fig:rhosh}). Because the cluster-centric radius of the radial velocity minimum  decreases  as redshift increases (see Sect. \ref{subsec:vrad} and Table \ref{table:rng_inf}), the infalling shells are at different radii. The infalling shell volume decreases for shells  nearer to the cluster center. Furthermore  the shell thickness is also not constant. Taken together these effects produce an increase of a factor of $\sim 2.0 $ in $M_{\rm shell}$ over the redshift range  we probe.

The slightly different mass distributions that characterize the cluster samples as a function of redshift (Table \ref{table:3dinfo}) also affect the redshift dependence of $M_{\rm shell}$. The bottom panel of Fig. \ref{fig:rhosh} shows the ratios of the median cumulative mass profiles.
For the highest redshift clusters, the median mass profiles are $\sim 20\%$ below the median profile at $z=0.01$ decreasing the correlation between $M_{\rm shell}$ and redshift by a comparable fraction. This effect couples with the effects of the relative density profiles and produces a combined factor of $\sim 1.6$ increase in $M_{\rm shell}$ over the range sampled by IllustrisTNG. In other words, the cluster mass distribution and the relative profiles as a function of redshift account for the lack of dependence of $M_{\rm shell}$ on redshift in the sample of clusters simulated with IllustrisTNG.

\end{appendix}

\end{document}